\documentclass[a4paper,fleqn]{cas-sc}

\usepackage[authoryear,longnamesfirst]{natbib}
\usepackage{graphicx}%
\usepackage{multirow}%
\usepackage{amsmath,amssymb,amsfonts}%
\usepackage{amsthm}%
\usepackage{mathrsfs}%
\usepackage[title]{appendix}%
\usepackage{xcolor}%
\usepackage{textcomp}%
\usepackage{manyfoot}%
\usepackage{booktabs}%
\usepackage{algorithm}%
\usepackage{algorithmicx}%
\usepackage{algpseudocode}%
\usepackage{listings}%
\usepackage{tikz}%
\usetikzlibrary{arrows.meta,positioning}%
\def\tsc#1{\csdef{#1}{\textsc{\lowercase{#1}}\xspace}}
\tsc{WGM}
\tsc{QE}

\usepackage{array,longtable,multirow,booktabs}
\newcolumntype{L}[1]{>{\raggedright\arraybackslash}p{#1}}
\renewcommand{\arraystretch}{1.12}
\usepackage{xcolor}
\newif\ifrevisions
\revisionstrue

\ifrevisions
  
  \newcommand{\new}[1]{{\color{black}#1}}
\else
  
  \newcommand{\new}[1]{#1}
\fi

\usepackage{enumitem}
\usepackage[most]{tcolorbox}

\setlist[itemize]{leftmargin=*, itemsep=2pt, topsep=2pt}
\setlist[enumerate]{leftmargin=*, itemsep=2pt, topsep=2pt}

\newtcolorbox{vizbox}[2][]{scaffbox, title={#2}, #1}
\newcommand{\descpt}[1]{\textit{Description (to participants):} #1}
\newcommand{\guiding}{\textbf{Guiding Questions:}}
\newcommand{\knowledge}{\textbf{Knowledge to convey:}}

\tcbset{
  scaffbox/.style={
    colback=white, colframe=black!60, boxrule=.4pt,
    arc=2mm, left=6pt, right=6pt, top=6pt, bottom=6pt,
    fonttitle=\bfseries, coltitle=black,
    colbacktitle=white,                 
    boxed title style={frame empty, interior empty, boxrule=0pt},
    toptitle=0pt, bottomtitle=2pt,      
    before skip=8pt, after skip=10pt
  }
}

\begin{document}
\let\WriteBookmarks\relax

\shorttitle{The Agency Gap in AI-Supported Writing}    

\shortauthors{Jin et al.}  

\title [mode = title]{The Agency Gap in AI-Supported Writing: How Reactive and Proactive Agent Designs Shape Multimodal Reasoning}  

\author[]{Yueqiao Jin}

\author[]{Kaixun Yang}

\author[]{Roberto Martinez-Maldonado}

\author[]{Dragan Gašević}\cormark[1]
\ead{dgasevic@hku.hk}

\author[]{Lixiang Yan}\cormark[1]
\ead{lixiangyan@tsinghua.edu.cn}

\cortext[cor1]{Corresponding author}

\affiliation[1]{
  organization={Centre for Learning Analytics at Monash, Faculty of Information Technology, Monash University},
  city={Clayton},
  state={Victoria},
  country={Australia}
}

\affiliation[2]{
  organization={Faculty of Education and School of Computing \& Data Science, The University of Hong Kong},
  city={Hong Kong}, 
  country={SAR People's Republic of China}
}

\affiliation[3]{
  organization={School of Education, Tsinghua University},
  city={Beijing},
  country={People's Republic of China}
}

\begin{abstract}
\new{Generative AI is becoming part of academic writing, but its educational value depends on how control is shared between learner and system. This study examined an \textit{agency gap}: performance differences that may arise when AI agent initiative is misaligned with learners' generative AI literacy. \new{Seventy-nine medical and nursing students completed two multimodal analytical writing tasks using healthcare simulation data visualisations. They were randomly assigned to a reactive agent that responded only when prompted or a proactive agent that provided sequenced questions and feedback. Generative AI literacy was measured using the validated 20-item Generative AI Literacy Assessment Test.} Epistemic network analysis showed that proactive interaction created stronger links among conceptual reasoning, evidence use, and constructive engagement, whereas reactive interaction was more factual and procedural. Ordinal regression showed that generative AI literacy predicted immediate independent writing performance after support was removed, particularly for visual data integration, critical thinking, and overall quality. \new{Condition-specific mediation estimates showed a literacy--performance association in the reactive condition but not in the proactive condition; however, the indirect effects and literacy-by-design interactions were not significant. This pattern is consistent with smaller literacy-related performance differences under proactive scaffolding, but it does not establish a compensatory causal effect.} Learner reflections indicated that effective AI writing support requires contextual feedback, dialogic scaffolding, and calibration of initiative to learner needs and task complexity. These findings position interaction design as a potential mechanism for supporting equitable and agency-supportive educational AI agents.}
\end{abstract}


\maketitle

\section{Introduction}\label{sec1}

Generative artificial intelligence (GenAI) is rapidly transforming how students learn, create, and communicate in higher education. Tools such as ChatGPT and Gemini now act as conversational collaborators capable of generating text, interpreting visuals, and providing feedback across academic disciplines \citep{kim2025exploring, nguyen2024human, Belkina_2025}. This development has shifted the focus of educational technology research from algorithmic performance to the quality of human-AI interaction, how learners engage with, respond to, and learn from AI agents \citep{Akpan_2024, bozkurt_2023, Song_2025}. These interactions are not neutral: they influence students’ motivation, self-efficacy, and agency in ways that can either extend or constrain meaningful learning \citep{bozkurt_2024, Wu_2025, pan_2025}. In academic writing, a cornerstone of higher education, GenAI has introduced new opportunities for scaffolding reasoning and expression, yet it also raises a central question: do such tools cultivate independent writing competence, or do they risk creating dependency on AI-generated guidance? \citep{zhai2024effects, yang2025modifying, wang2024exploring}  

A key determinant of these outcomes lies in the interaction design. AI agents can be configured as \textit{reactive}, responding only when prompted, or \textit{proactive}, taking the initiative to scaffold understanding through questions and feedback \citep{park2023generative, lee2024design}. Reactive designs foreground student autonomy but require prompting skill and metacognitive awareness, while proactive designs reduce cognitive demand by steering engagement, yet may over-structure learning if applied too rigidly \citep{yan2024vizchat, bai_2024, Yan_2025}. Research on conversational support in educational technology indicates that the nature of interaction design can shape learner engagement, agency, and feedback uptake in both positive and unintended ways \citep{oertel2020engagement, Looi_2025, an2025impacts}. Despite the growing integration of AI agents into higher education, empirical evidence remains scarce regarding how these different interaction designs influence students’ reasoning, feedback use, and learning transfer, particularly in multimodal academic writing, where visual interpretation and synthesis are central to meaning-making \citep{ye2024generative, fernandez2024data, milesi2024data}.  

Beyond AI agent interaction design, the learner’s own GenAI literacy, knowledge of how generative agents function, skill in prompting and evaluating outputs, and awareness of ethical and contextual limitations, plays a decisive role in determining the success of human-AI collaboration \citep{jin2024glat, annapureddy2024generative, bozkurt_2024}. Recent frameworks conceptualise GenAI literacy as a multidimensional construct encompassing technical understanding, evaluative judgment, and critical awareness of the social implications of AI tools \citep{stolpe_2024, Ng_2021, celik2023exploring}. Students with higher GenAI literacy often use AI strategically to enhance conceptual understanding, sustain learning outcomes, and transfer knowledge beyond the mediated environment \citep{fan2025beware, yang2025modifying, Wu_2025}. Conversely, those with lower literacy may struggle to engage effectively without continuous scaffolding and may be more susceptible to overreliance or superficial engagement with AI-generated outputs \citep{stadler2024cognitive, zhai2024effects, abbas2024harmful}.  

\new{This study addresses these gaps by examining how interaction design (reactive vs.\ proactive) and GenAI literacy jointly shape multimodal analytical writing. We use the term \textit{agency gap} as a relational conceptualisation of a mismatch between the initiative demanded by an AI agent (e.g., designed to be reactive or proactive) and the learner's capacity to initiate, monitor, evaluate, and internalise AI-supported reasoning. The concept therefore concerns the alignment of learner capability and AI agent interaction design rather than an individual deficit or a fixed property of either party. By analysing student--AI dialogue, writing performance, and reflective accounts, the study examines observable process and outcome patterns that are consistent with this mismatch \citep{chi2014icap, fan2025beware, xie2024can}. The study's primary theoretical contribution is to identify the distribution of initiative as a design-level boundary condition, specifying when and for whom AI interaction designs may support or constrain learner agency. It connects three levels that are often studied separately: AI agent initiative, learner GenAI literacy, and the organisation of epistemic work during and immediately after AI-supported writing. This framing advances prior work by asking not only whether AI support improves an outcome, but also for whom a particular initiative structure creates opportunities to exercise agency, engage in reasoning, and perform independently once support is removed.}

\section{Background}\label{sec2}

\subsection{Human-AI Interaction in Academic Writing}

Academic writing is a high-level cognitive activity that requires students to synthesise ideas, interpret evidence, and construct coherent arguments. It demands self-regulation, reflection, and the capacity to translate conceptual understanding into structured expression \citep{lea2006academic, wingate2012argument, kuhn2005education}. Within higher education, writing is not only a means of assessment but also a process through which disciplinary knowledge is built, negotiated, and communicated \citep{swales1994academic, paltridge2004academic, hyland2021teaching}. Students must navigate between abstract reasoning and concrete representation, often integrating multimodal information, such as data charts, network diagrams, or visual analytics, into their textual arguments \citep{jewitt2006technology, fernandez2021storytelling, ye2024generative}. Mastery of such multimodal analytical writing represents a synthesis of visual literacy, critical thinking, and rhetorical awareness \citep{felten2008visual, facione2000disposition, arneson2018visual}.

The emergence of GenAI has reshaped this process in both practice and pedagogy. AI-driven writing agents now act as dynamic collaborators that can analyse visuals, generate textual summaries, and scaffold students’ reasoning through conversational interaction \citep{lee2022coauthor, nguyen2024human, kim2025exploring}. Rather than functioning as static tools, these AI agents mediate cognitive and linguistic processes by offering feedback, language models, and adaptive prompts that resemble elements of peer review and mentoring \citep{lee2024design, lee2022coauthor, cardon2023challenges}. This shift has prompted renewed attention to the \textit{interaction process}, how students communicate with AI agents, regulate their thought processes through dialogue, and integrate generated content into authentic writing tasks \citep{yang2025modifying, shibani2023visual, Akpan_2024}. 

Research on human-AI collaboration in educational settings suggests that effective learning derives less from model accuracy than from the \textit{quality of engagement} achieved during interaction \citep{yang2024ink, Wu_2025, fan2025beware}. Students who iteratively prompt, critique, and refine AI-generated outputs often demonstrate more metacognitive awareness and deeper reasoning than those who adopt AI responses uncritically \citep{yan2023impact, zhai2024effects, yang2025modifying}. This aligns with evidence from studies of human-AI agent interaction showing that active participation and reflection mediate the benefits of intelligent feedback agents \citep{oertel2020engagement, bai_2024, pan_2025}. However, most existing research still privileges surface metrics, such as perceived writing quality, productivity, or satisfaction, while neglecting the underlying socio-cognitive mechanisms that drive effective co-creation between humans and AI agents \citep{zhai2024effects, jin2025chatting, li2023can}.

Despite this growing body of work, substantial gaps remain regarding how specific AI interaction designs influence learning dynamics. In particular, few empirical studies have compared the consequences of \textit{reactive} versus \textit{proactive} AI agent designs for students’ reasoning, metacognition, and sense of authorship \citep{yan2024vizchat, park2023generative, hou2024effects}. As the boundaries between learner, tool, and feedback source continue to blur, understanding how interaction design shapes the depth of reasoning, self-regulation, and long-term knowledge transfer becomes critical for developing evidence-based frameworks for AI-assisted writing in higher education \citep{Yan_2025, Ng_2021, Belkina_2025}. \new{Accordingly, the present study shifts the unit of explanation from the quality of an AI output to the organisation of epistemic processes within the interaction: who initiates inquiry, who monitors progress, and how observations are connected to concepts and evidence.}

\subsection{Reactive and Proactive AI Interaction Designs}

In the context of this paper, the concept of \textit{interaction design} refers to how the initiative, timing, and depth of AI responses are structured to support the learner. Regarding AI-assisted writing, two contrasting paradigms can be identified: \textit{reactive} and \textit{proactive} interaction designs \citep{jin2025chatting}. These modes differ not only in conversational dynamics but also in how they distribute cognitive responsibility and agency between the learner and the AI agent \citep{park2023generative, lee2024design, yan2024vizchat}.  

Reactive AI agents are designed to adopt a low-initiative stance: they wait for the learner’s input and respond only when prompted. This design positions the student as the principal driver of interaction, encouraging autonomy and active regulation of the learning process \citep{Ait_Baha_2023, Song_2025, Wu_2025}. Empirical research suggests that reactive agents can foster deeper engagement and critical reflection when learners demonstrate strong metacognitive awareness and the ability to formulate purposeful prompts \citep{nguyen2024human, yang2025modifying, fan2025beware}. They promote ownership of learning by requiring students to define rhetorical goals, articulate uncertainties, and assess the adequacy of AI-generated responses. However, this approach presupposes a certain level of GenAI literacy and confidence in navigating conversational parameters. Learners with limited prompt literacy or lower technological self-efficacy may struggle to elicit meaningful feedback, resulting in surface-level revisions or unproductive cognitive load \citep{Ng_2024, bozkurt_2024, annapureddy2024generative}. Consequently, while reactive AI agents can strengthen self-direction among experienced users, they risk amplifying existing inequities by privileging those already equipped with advanced digital and metacognitive competencies \citep{jin2024glat, celik2023exploring, Belkina_2025}.  

Proactive AI agents, by contrast, are designed to draw on principles of guided scaffolding and formative assessment to take a more directive role in supporting the learner’s cognitive and linguistic process. Grounded in Vygotskian theories of mediated learning and the zone of proximal development \citep{vygotsky1978mind, gibbons2002scaffolding}, proactive AI agents initiate questions, propose directions, and strategically sequence tasks to sustain learner engagement. This approach mirrors tutor-led feedback cycles that diagnose gaps, prompt reflection, and model expert reasoning without completely removing the learner’s agency \citep{hattie2007power, pan_2025, bai_2024}. Empirical studies show that proactive AI scaffolding can improve comprehension of complex materials, sustain self-regulated learning, and enhance reasoning depth during ill-structured or analytical writing tasks \citep{Yan_2025, Looi_2025, hou2024effects}. Through adaptive dialogue and timely intervention, proactive designs help reduce cognitive load and maintain motivation, particularly for novices or students who encounter conceptual uncertainty \citep{Lyu_2024, Akpan_2024, an2025impacts}.  

Nevertheless, proactive designs carry notable trade-offs. When the AI’s feedback becomes overly prescriptive, it may inadvertently constrain exploration, diminish the learner’s agency, and encourage overreliance on machine-generated direction \citep{stadler2024cognitive, zhai2024effects, Wu_2025}. Evidence from studies on scaffolding and feedback also indicates that too much structure can restrict students’ capacity for independent reasoning and limit the desirable difficulty needed for durable learning \citep{nicholas2015learning, chi2014icap, bjork1994testing}. Balancing guidance and autonomy therefore remains a central pedagogical and ethical design challenge. Understanding when to give initiative to the learner and when to prompt with supportive interventions is essential for developing AI agents that enhance learning rather than replace it. Future comparative research should investigate how reactive and proactive modes differentially affect engagement patterns, cognitive strategies, and knowledge transfer in authentic higher education contexts such as multimodal analytical writing \citep{yan2024genai, jin2025chatting, Belkina_2025}. \new{In this study, reactive and proactive AI agent designs are therefore treated as different allocations of conversational initiative rather than as simple low- and high-support conditions. This distinction allows the analysis to separate the amount of assistance from the locus of regulation and to examine whether a fixed level of initiative fits learners with different capabilities.}

\subsection{Generative AI Literacy and Learner Agency}

Interaction design alone cannot fully account for the wide variation in learning outcomes; equally important is the learner’s competence in navigating and utilising GenAI tools effectively. This competence, termed \textit{generative AI literacy}, extends beyond general digital or information literacy to encompass conceptual understanding, procedural skill, and critical awareness of generative technologies \citep{jin2024glat, bozkurt_2024, Ng_2021}. It involves understanding how large language models are trained and fine-tuned, recognising the probabilistic nature of their outputs, crafting effective prompts to elicit accurate responses, and evaluating those responses with an awareness of ethical, epistemological, and contextual implications \citep{Zhao_2024, annapureddy2024generative, stolpe_2024}. In this way, GenAI literacy integrates cognitive, technical, and socio-ethical dimensions that shape how students perceive, interpret, and employ AI as a learning collaborator rather than merely a knowledge source \citep{ooi2023potential, celik2023exploring, Akpan_2024}.

Students with high levels of GenAI literacy are more likely to approach AI tools strategically, using them to test conceptual hypotheses, seek clarification, explore alternative framings, and refine arguments grounded in their own reasoning processes \citep{yan2024vizchat, yang2025modifying, Wu_2025}. They engage in reflective prompting, critiquing, rewording, and validating AI suggestions against disciplinary standards, behaviours associated with higher-order metacognition and agency \citep{fan2025beware, li2023can, yang2024ink}. In contrast, learners with lower literacy may rely passively on AI-generated text, misinterpret probabilistic outputs as definitive answers, or withdraw when AI feedback conflicts with prior understanding \citep{chiu_2024, zhai2024effects, abbas2024harmful}. Such disparities mirror broader findings that technological proficiency amplifies differential learning gains in digital environments, creating what is sometimes termed an “agency gap” between adaptive and dependent users \citep{bozkurt_2024, jin2024glat, Belkina_2025}.

\new{GenAI literacy may also support learners’ capacity to apply AI-supported reasoning when assistance is reduced or removed. Students with stronger literacy may be better positioned to evaluate and internalise useful reasoning strategies, whereas students with weaker literacy may find independent application more difficult \citep{xie2024can, Ng_2024, Yan_2025, stadler2024cognitive, zhai2024effects, an2025impacts}. However, the present study assesses performance only on an immediate task completed after agent removal, which examines near-term independent application rather than durable learning transfer. This distinction positions GenAI literacy as a potentially important instructional target while avoiding claims about retention or transfer that require delayed assessment.}

Despite emerging conceptual frameworks and measurement instruments such as the Generative AI Literacy Assessment Test (\textit{GLAT}) \citep{jin2024glat}, the dynamic interplay between literacy and AI agent design remains underexplored. Most current research treats AI literacy as a static trait, overlooking how learners’ competences evolve through interaction with varying pedagogical affordances of reactive and proactive agent designs \citep{bozkurt_2024, Ng_2021, yan2024genai}. Understanding how literacy modulates the cognitive and motivational impact of these designs is essential for tailoring GenAI-integrated educational agents that promote both equity and productive agency in learners. \new{Within the framework, GenAI literacy is not synonymous with learner agency. Rather, it is a capability that may enable agency to be exercised effectively when the interaction design requires the learner to formulate requests, evaluate outputs, and decide how AI contributions should enter the writing process.}

\subsection{Interaction Design, Equity, and the “Agency Gap”}

\new{Interaction design and GenAI literacy jointly shape how agency is distributed within human--AI partnerships, extending existing accounts of learner control and regulation beyond purely human learning contexts \citep{Bandura2001,nopas2025algorithmic}. Learner agency refers to the capacity to act intentionally and make informed choices, whereas self-regulated learning describes the processes through which learners set goals, monitor progress, select strategies, and evaluate outcomes \citep{Panadero_2017}. Co-regulation and shared regulation further extend this perspective by describing how regulatory responsibility can be supported by, or distributed across, interaction partners \citep{Hadwin_2018}. Adaptive scaffolding provides an instructional mechanism for adjusting support to learner needs \citep{van2010scaffolding}, while the expertise reversal effect suggests that guidance beneficial for less experienced learners may become redundant or constraining as expertise increases \citep{kalyuga2009expertise}.}

\new{The \textit{agency gap} construct captures a specific design--learner mismatch in AI-mediated activity rather than replacing established theories of agency, regulation, or scaffolding \citep{Bandura2001}. It describes situations in which an AI agent requires more self-initiation and evaluative control than a learner can currently exercise, or alternatively, where excessive AI agent initiative reduces opportunities for a capable learner to direct and test their own reasoning \citep{Bandura2001,Kim2026}. The former represents \textit{under-support}, such as when learners with lower GenAI literacy interact with strongly reactive environments; the latter represents \textit{over-direction}, such as when highly capable learners engage with rigidly proactive environments \citep{Kim2026,Pozdniakov2023ectel}. Because task complexity and disciplinary expectations shape the demands placed on learners, neither agent design is assumed to be universally superior.}

\begin{figure}[!htbp]
    \centering
    \includegraphics[width=0.85\linewidth]{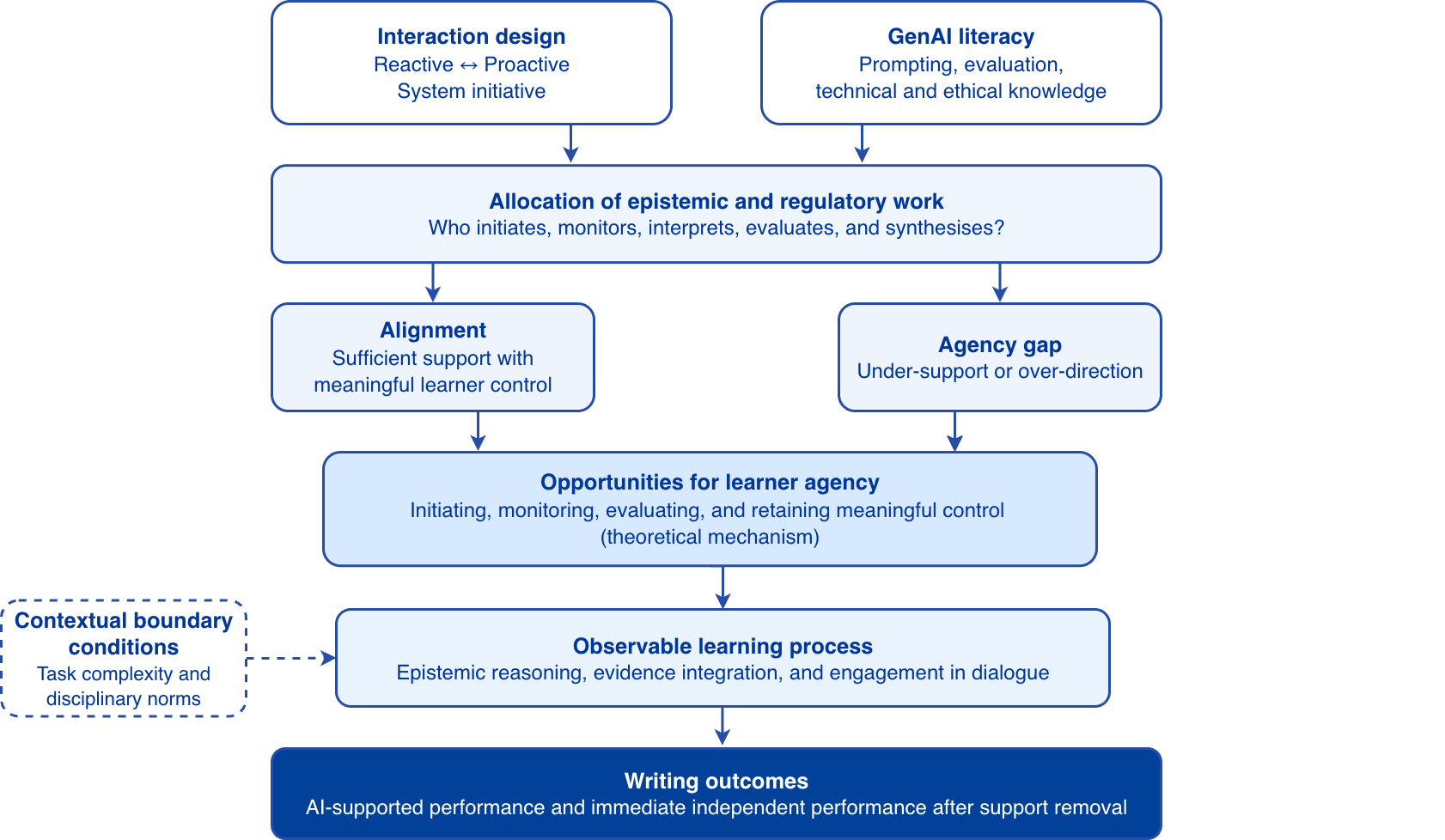}
    \caption{\new{Relational model of the agency gap in AI-supported writing. Interaction design and GenAI literacy are theorised to shape how epistemic and regulatory work is allocated between learner and system. Alignment and misalignment are expected to affect opportunities for learner agency and, in turn, the organisation of dialogue and immediate writing outcomes. Learner agency is represented as a theoretical mechanism and was not measured as a standalone scale in this study.}}
    \label{fig:agency-gap-model}
\end{figure}

\new{The agency gap perspective proposes that alignment between AI initiative and learner capability influences how epistemic work is distributed during human--AI interaction \citep{Bandura2001,Kim2026}. Specifically, interaction design and GenAI literacy may jointly shape who initiates questions, interprets evidence, monitors understanding, and evaluates feedback, with consequences for dialogue structures and immediate independent writing performance \citep{Kim2026}. Alignment should preserve meaningful learner control while providing sufficient support, whereas misalignment may create unequal opportunities for learners to exercise agency \citep{Bandura2001,Kim2026}. The present study does not conceptualise the agency gap as an established or directly measured latent construct; instead, it examines this emerging conceptualisation through converging evidence from discourse networks, writing outcomes, and learner reflections. Figure~\ref{fig:agency-gap-model} summarises the framework tested in the study by positioning agent initiative and GenAI literacy as inputs, the allocation of epistemic work and opportunities for learner agency as mechanisms, and reasoning, engagement, and writing performance as observable consequences. Thus, the contribution is not a new general theory of agency, but a testable account of how human--AI design alignment may shape participation and performance.}

\subsection{Research Questions}

To advance this line of inquiry, the present study investigates how interaction design and GenAI literacy jointly shape learning processes and outcomes in multimodal analytical writing. Specifically, we examine (1) how different human-AI interaction designs influence epistemic reasoning and engagement, (2) how GenAI literacy predicts performance during and after AI-assisted writing, and (3) how learners experience these designs as supportive or constraining. Through an integrated mixed-method approach combining discourse analysis, regression and mediation modelling, and thematic analysis of learner reflections, the study addresses the following research questions:

\begin{itemize}
    \item \textbf{RQ1:} How do reactive and proactive human-AI interaction designs differ in the network structure of epistemic reasoning and learner engagement during multimodal analytical writing?

    \item \textbf{RQ2:} To what extent does GenAI literacy predict learning outcomes under different interaction designs (reactive vs.\ proactive), both during AI-supported writing and after AI is withdrawn?

    \item \textbf{RQ3:} Does GenAI literacy exert its influence directly on independent writing performance, or indirectly through AI-supported task performance, and how is this relationship moderated by interaction design?

    \item \textbf{RQ4:} What interaction design heuristics can be derived from learners’ experiences with reactive and proactive AI agents to guide the development of equitable and autonomy-supportive multimodal writing systems?
\end{itemize}

Together, these questions seek to bridge the conceptual and empirical divide between human-AI interaction research and educational design. By analysing how design features and learner competencies jointly shape both process and outcome, the study aims to generate actionable insights for developing interaction models that promote meaningful engagement, equitable support, and sustainable learning in higher education.

\section{Methods}

\subsection{Learning Contexts}
The learning context focused on writing based on visual information, a critical component of academic literacy and analytical communication. The task was modelled after visual interpretation exercises such as the IELTS Academic Writing Task~1 \citep{ielts2025writing}. Participants were presented with three distinct visual representations, a bar chart, a network diagram, and a ward map (described in detail in Section~\ref{sec-materials}), and were instructed to compose a short analytical response of approximately 150 words. Each task prompt asked them to interpret the visuals and address a question about team performance (e.g., \textit{"How did the two nurses manage the primary patient (Amy) while attending to other beds, focusing on their task prioritisation, verbal communication, and stress levels?"}). The visuals were generated from authentic data collected in healthcare simulation scenarios, ensuring realism while preventing prior familiarity effects. Consequently, participants’ understanding and interpretation relied solely on the information provided within the task. All activities were conducted through a purpose-built online platform (see Section~\ref{sec-platform}).

\subsection{Participants}

\new{A total of 81 university students participated in the study. Two were excluded from analysis because their written responses contained fewer than 20 words, resulting in a final sample of 79 participants. Recruitment was conducted via Prolific Academic, and participants were compensated £8 for approximately one hour of engagement. An \textit{a priori} power analysis for the primary two-condition comparison indicated a minimum of 72 participants (36 per condition) for a medium effect ($f = 0.25$), power of 0.80, and $\alpha = 0.05$. This calculation was used to plan the primary comparison and should not be interpreted as guaranteeing adequate power for every secondary model. The ENA and ordinal regression analyses therefore used prespecified, parsimonious models together with nonparametric tests, effect sizes, and diagnostic checks. The condition-specific mediation analyses involved 39 and 40 participants, respectively, and were treated as exploratory; their uncertainty was evaluated using 5{,}000-resample bootstrap confidence intervals rather than significance alone.}

\new{Participants were randomly assigned to the reactive agent ($n = 39$) or proactive agent ($n = 40$) condition. All were current students or recent graduates from medical or nursing programs. The sample comprised 45 females and 34 males, primarily aged 18--24 ($n = 28$) and 25--34 ($n = 31$). Geographically, participants were distributed across North/Central America ($n = 34$), Europe ($n = 21$), Africa ($n = 15$), Australia ($n = 4$), South America ($n = 2$), and other regions ($n = 3$). Educational backgrounds included Bachelor's degrees ($n = 36$), high school diplomas or equivalents ($n = 18$), Master's degrees ($n = 11$), vocational diplomas or training ($n = 7$), Doctoral degrees ($n = 4$), and other qualifications ($n = 3$). Familiarity with GenAI tools was reported as intermediate ($n = 42$), beginner ($n = 25$), advanced ($n = 9$), expert ($n = 2$), and none ($n = 1$).}

\subsection{Materials and Apparatus}
\subsubsection{Learning Materials}
\label{sec-materials}

\begin{figure}
    \centering
    \includegraphics[width=1\linewidth]{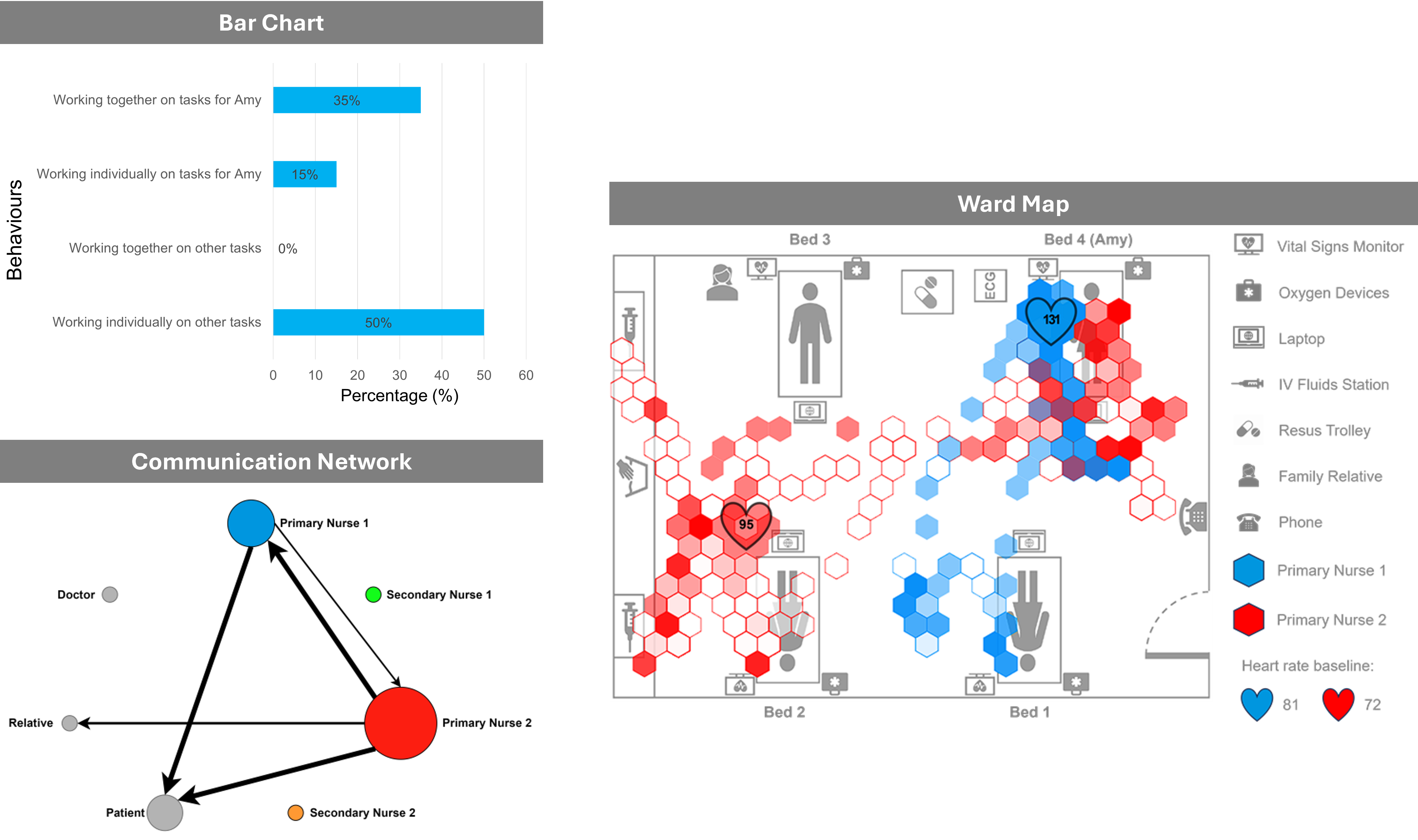}
    \caption{Three visuals showing (i) team behaviours (bar chart), (ii) team communication patterns (network diagram), and (iii) movement, communication intensity, and physiological responses (ward heatmap).}
    \label{fig-visuals}
\end{figure}

The three visual representations used in the learning tasks were generated from multimodal data collected during a high-fidelity healthcare simulation, where teams of four nursing students (two primary and two secondary nurses) managed acute clinical scenarios in a realistic ward environment. The simulation aimed to develop prioritisation, communication, and teamwork skills, with data captured from multiple sources: positional tracking (x--y coordinates and body orientation), audio interactions via wireless headsets, and physiological measures (heart rate) using Fitbit Sense wristbands. These data produced three progressively complex visualisations (Figure~\ref{fig-visuals}): (i) a \textit{bar chart} showing time spent on different behaviours, reflecting team-level resource allocation and prioritisation; (ii) a \textit{communication network} combining positional and audio data to visualise the frequency and direction of verbal interactions among team members; and (iii) a \textit{ward map} integrating positional, audio, and physiological data through heatmap techniques to depict communication intensity, spatial movement, and areas of peak physiological arousal \citep{goldsberry2012courtvision}.

\subsubsection{Research Platform}
\label{sec-platform}
\begin{figure}
    \centering
    \includegraphics[width=1\linewidth]{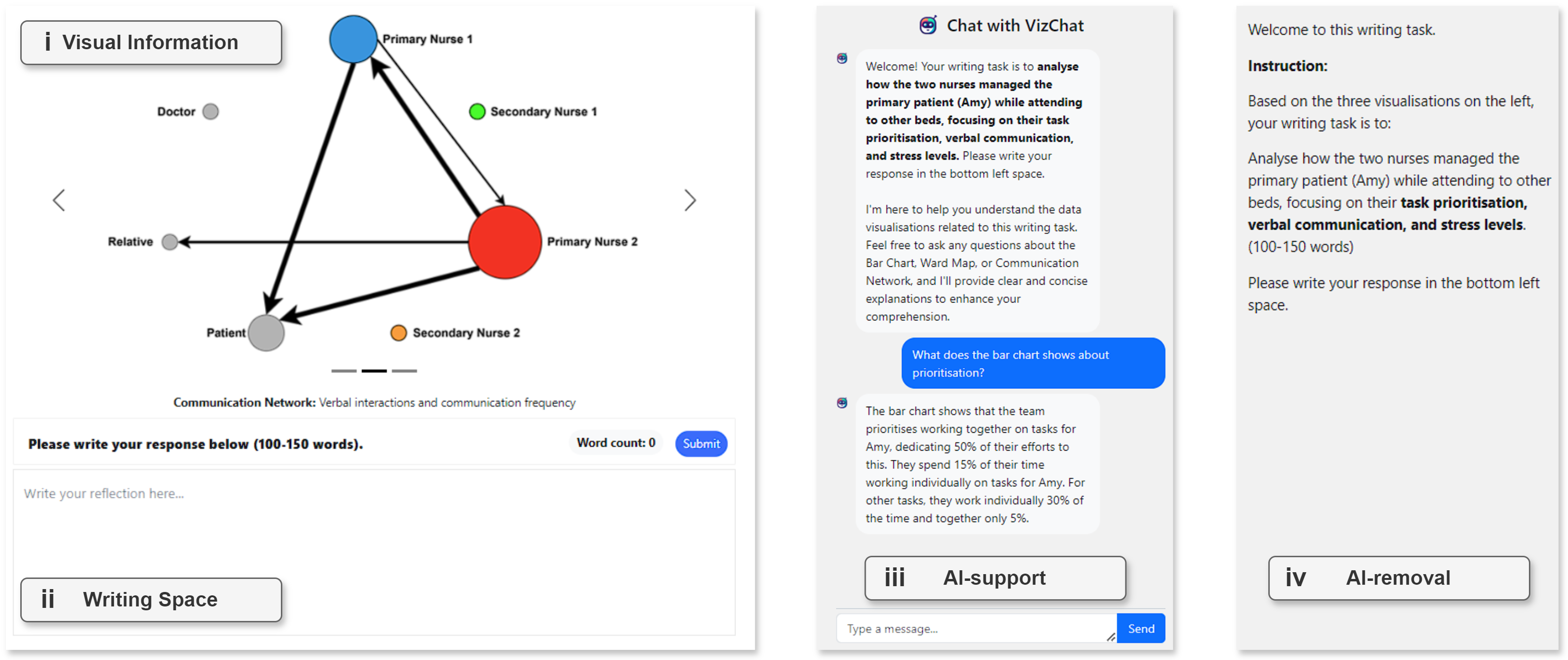}
    \caption{Research platform showing (i) visual information display, (ii) academic writing space, and (iii) either AI-supported chat function (reactive or proactive) or (iv) standard task instruction (AI-removal condition).}
    \label{fig-platform}
\end{figure}

The academic writing tasks were conducted using a custom-built online platform comprising three components. The first component, the \textit{display area} (Figure~\ref{fig-platform}), presented one visual analytic at a time to minimise visual clutter and prevent information overload \citep{ellis2007taxonomy}. Participants could navigate between the three visuals using arrows on either side of the display. The second component, the \textit{writing space}, provided an area for composing analytical responses by synthesising information from the visuals. The third component varied by condition: in the AI-supported tasks, participants accessed a \textit{chat interface} offering real-time support from AI agents, which responded either reactively or proactively; in the non-AI condition, this component was replaced by static task instructions. The overall platform was intentionally minimalist \citep{stone2005user}, allowing participants to focus on interpreting the visual information and completing the writing tasks without distraction from unnecessary interface complexity.

\subsubsection{Agent Interaction Design}
\begin{figure}
    \centering
    \includegraphics[width=1\linewidth]{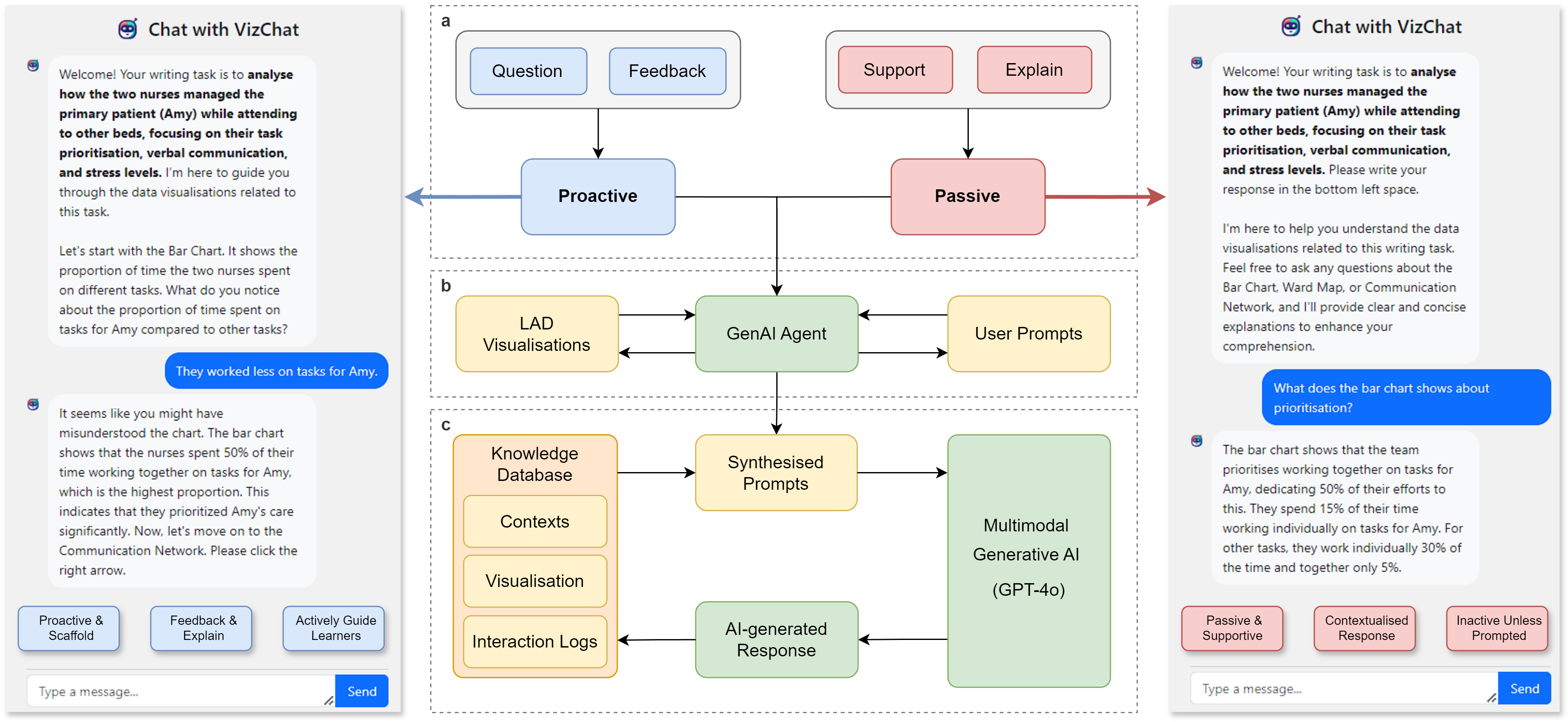}
    \caption{AI agent interaction design contrasting a reactive agent (red; right) responding only upon learner query and a proactive agent (blue; left) scaffolding learning through structured questioning and iterative feedback.}
    \label{figure-conditions}
\end{figure}

\new{Two AI agents were developed, \textit{reactive} and \textit{proactive}, to manipulate conversational initiative while holding other AI agent features constant (Figure~\ref{figure-conditions}). Both conditions used the same online interface, visual inputs, task instructions, VizChat architecture, Retrieval-Augmented Generation pipeline, visualisation knowledge base, and multimodal GPT-4o model \citep{yan2024vizchat, gao2023retrieval}. The intended experimental contrast was therefore not model capability or access to information, but whether the learner or the agent initiated and sequenced the support. Both agents began with the same welcome and task-orientation message.}

\new{In the reactive condition, generation was triggered only by a learner message. The agent answered the request using the current visual and retrieved task context, but it did not introduce an unsolicited question, initiate a follow-up sequence, or determine the next visual to inspect \citep{ma2023demonstration, yan2024vizchat}. Responsibility for identifying uncertainty, formulating a useful prompt, deciding whether further help was needed, and moving between visuals therefore remained with the learner.}

\new{In the proactive condition, intervention was governed by an expert-authored scaffold represented as a simple conversational state sequence rather than by a numerical learner model. The state variables were (a) the visual currently displayed and (b) which guiding questions for that visual had already been covered. When a participant entered a visual, the agent initiated a one-sentence orientation and the next ordered guiding question. After each learner response, it generated task- or process-level feedback with reference to the corresponding expert-authored ``knowledge to convey'' statement: it confirmed an interpretation that was aligned with the visual evidence, clarified a missing visual cue or relationship, or corrected an interpretation that contradicted the displayed evidence. It then advanced to the next non-repeated question. The prompt prohibited asking about a visual that was not currently displayed and prohibited repeating the same guiding question. After the prescribed questions for a visual were completed, the agent prompted the participant to move to the next visual; after all visuals were completed, it issued a final synthesis prompt for the written paragraph. Thus, the \textit{when} of intervention was determined by visual entry and completion of each learner turn, while the \textit{what} was determined by a fixed sequence of descriptions, guiding questions, target knowledge, and transition instructions (Appendix~\ref{sec:scaffold}).}

\begin{table}[!htbp]
\caption{\new{Proactive agent intervention logic and prompt structure}}
\label{tab:proactive-logic}
\centering
\new{\begin{tabular}{p{2.5cm}p{3.2cm}p{7.0cm}}
\hline
\textbf{Interaction state} & \textbf{Trigger/criterion} & \textbf{Agent action and design purpose} \\
\hline
Visual entry & A new visualisation is displayed & Provide a one-sentence description and initiate the first unanswered guiding question to orient attention to task-relevant features. \\
Learner response & The learner submits a response to the current question & Use the expert ``knowledge to convey'' statement to generate confirmatory, clarifying, or corrective feedback grounded in the visual evidence; do not compose the learner's final paragraph. \\
Within-visual progression & The current question has been addressed & Advance to the next unanswered question for that visual. Each question is asked once, and questions about non-displayed visuals are withheld. \\
Between-visual progression & All questions for the current visual are completed & Ask the learner to navigate to the next visual. The sequence contained one question for the bar chart, two for the communication network, and three for the ward map. \\
Task completion & All three visual sequences are complete & Prompt the learner to synthesise task prioritisation, communication, and stress evidence in a cohesive analytical paragraph. \\
\hline
\end{tabular}}
\end{table}

\new{The proactive prompt was designed from Hattie and Timperley's task- and process-feedback principles \citep{hattie2007power} and a conceptual approach to visual interpretation \citep{xun2004conceptual}. The sequence moved from noticing a visual feature, to interpreting its meaning, to connecting evidence across visuals, and finally to synthesis. Generated wording could vary with the learner's response, but the triggering events, question order, target knowledge, and completion rules were fixed across participants.}

\subsection{Measures and Instruments}

\subsubsection{Generative AI Literacy Measurement (GLAT)}
\new{Learners' generative AI literacy was measured using the Generative AI Literacy Assessment Test (GLAT; \citealt{jin2024glat}), a performance-based 20-item multiple-choice instrument. The items cover four dimensions represented in the full instrument, understanding, applying, creating with, and making ethical judgements about GenAI, and each item has four response options. A corrected total score was calculated, with higher scores indicating stronger GenAI literacy. The GLAT was selected instead of a self-report confidence scale because it assesses demonstrated knowledge and judgement relevant to prompting, evaluating, and using generative systems.}

\new{Published validation evidence includes expert review for face validity and a 204-participant study of item relevance, comprehensiveness, and clarity. Structural validity and internal consistency were examined using classical test theory and item response theory, supporting a stable two-parameter logistic model (Cronbach's $\alpha = 0.76$; $\chi^2(133) = 135.06$, $p > .05$; $\text{RMSEA} = 0.007$, 90\% CI [0, 0.035]). Because the 2PL model accounts for item difficulty and discrimination but not guessing, the present study applied the published correction-for-guessing formula \citep{thorndike1991measurement}:}
\begin{equation}
CS = R - \frac{W}{(C - 1)}
\end{equation}
\new{where $CS$ is the corrected score, $R$ the number of correct responses, $W$ the number of incorrect responses, and $C$ the number of choices per item (four in the GLAT). The present analysis treated the corrected total as an observed continuous predictor and relied on the instrument's published validation evidence; no new factor validation was attempted in the current modest sample. The complete item set and dimension labels are provided in Appendix~\ref{sec:appendix-GLAT}.}

\subsubsection{Domain Knowledge Measurement (VLAT)}

Visualisation literacy, an essential component of domain knowledge for effective multimodal writing \citep{felten2008visual, lee2016vlat}, was measured using the Visualisation Literacy Assessment Test (mini-VLAT; \citealt{pandey2023mini}). This 12-item multiple-choice instrument offers validity and reliability comparable to the original 53-item VLAT \citep{lee2016vlat} while providing a concise format suitable for extended experimental tasks. Its brevity helped minimise cognitive load during assessment. Prior studies have confirmed the instrument’s structural validity and internal consistency \citep{pandey2023mini}. To control for potential guessing effects, participants’ scores were adjusted using the correction-for-guessing formula \citep{thorndike1991measurement}. The visualisation types and corresponding questions are detailed in Appendix~\ref{sec:appendix-vlat}.

\subsubsection{Academic Writing Assessment Rubric}

A comprehensive assessment rubric was developed to evaluate participants’ academic writing performance in tasks involving visual data analysis. Grounded in prior literature and established frameworks in academic writing research, five distinct evaluation criteria were identified (Table~\ref{tab:rubric}): \textit{Insightfulness}, \textit{Visual Data Integration}, \textit{Organisation and Coherence}, \textit{Linguistic Quality}, and \textit{Critical Thinking}. Each criterion was rated on a five-point Likert scale (Level~1, \textit{Poor} to Level~5, \textit{Excellent}), with detailed descriptors and exemplar responses provided in the full rubric (Appendix~\ref{sec:appendix-rubric}).

\begin{table}
\caption{Assessment Metrics for Academic Writing with Visual Data}
 \label{tab:rubric}
\centering
\begin{tabular}{p{2.5cm}p{10cm}p{1.5cm}}
\hline
\textbf{Metric} & \textbf{Description} & \textbf{Reliability (\( r_s\))}\\
\hline
Insightfulness &  
Assesses the originality, depth, and relevance of insights. High-quality responses identify subtle patterns, offer holistic interpretations, and provide conclusions that extend beyond obvious points. Low-quality responses are surface-level observations or simplistic interpretations \citep{kuhn2005education, chanock2000comments}. &
0.82\\
\\
Visual Data Integration&  
Measures the effectiveness and clarity with which visual data (charts, graphs, communication networks, heart rate information) is explicitly interpreted, combined, and connected into the written analysis. High-quality responses synthesize multiple visual data sources cohesively, while low-quality responses merely mention data without synthesizing or miss integrating data entirely \citep{jewitt2006technology}. &
0.85\\
\\
Organisation and Coherence &  
Evaluates the clarity, logical structure, and connectedness of ideas presented in the writing. High-quality responses have logical structure, smooth transitions between points, and clearly organised arguments. Low-quality responses lack clarity, exhibit abrupt transitions, or are disorganised \citep{paltridge2004academic}. &
0.75\\
\\
Linguistic Quality &  
Reflects grammar accuracy, vocabulary appropriateness, sentence fluency, and overall clarity of written expression. High-quality responses utilize clear, precise, and varied language with minimal errors, while low-quality responses contain grammatical errors, limited vocabulary, or unclear expression \citep{biber2010challenging}. &
0.89\\
\\
Critical Thinking &  
Evaluates the level of careful thought, complexity, consideration of alternative viewpoints, and strength of conclusions and recommendations. High-quality responses deeply analyze complexities and implications, considering broader contexts, contingencies, or solutions. Low-quality responses show limited or superficial analytical engagement and fail to address alternative perspectives or implications \citep{facione2000disposition, moon2007critical}. &
0.80\\
\hline
\end{tabular}
\end{table}

\subsubsection{Scoring Procedures and Inter-Rater Reliability}

\new{Two independent coders evaluated each participant's writing samples across the AI-support and AI-removal phases using the rubric in Appendix~\ref{sec:appendix-rubric}. Before formal scoring, the coders completed iterative training with 10 sample responses, compared criterion-level decisions, and discussed the interpretation of descriptors and exemplars until a common scoring protocol was established. All study essays were then rated independently without consultation during the initial scoring pass.}

\new{Inter-rater consistency was assessed using Spearman's rank correlations because each criterion was scored on an ordered five-level scale. Correlations were high across all dimensions ($r_s \geq .75$, $p < .001$; Table~\ref{tab:rubric}). Disagreements were handled by aggregation rather than by replacing one rater's judgement with a post hoc consensus score: the two independent criterion ratings were averaged to produce the final score for each dimension, and these dimension scores were then combined to form the overall \textit{Composite Score}. This procedure preserves information from both raters while reducing the influence of any single judgement. Exemplar excerpts for every scoring level are included in Appendix~\ref{sec:appendix-rubric} to support transparency and replication.}

\subsubsection{Epistemic Quality and ICAP Engagement}

Two complementary frameworks were applied to characterise the human-AI dialogues: the epistemic dimension of the argumentation knowledge construction framework \citep{weinberger_framework_2006} and the ICAP engagement framework \citep{chi2014icap}. The Epistemic Dimension captured the quality and correctness of reasoning expressed in each utterance, while the ICAP framework indicated the depth of cognitive engagement of human learners (Table \ref{tab:codebook}). All utterances were segmented at the message level. The Epistemic Dimension was applied to every utterance (human and AI), and the ICAP framework only to learner utterances. Two trained coders independently coded an initial random sample of 25\% of the data to establish reliability. After satisfactory agreement was reached, both coders coded the remaining 75\% of the corpus independently, with discrepancies resolved through discussion. In total, 554 utterances (310 AI and 244 human) were coded for epistemic quality, and 244 human utterances for engagement. Cohen’s $\kappa$ coefficients were calculated for each code (Table~\ref{tab:codebook}). Overall, inter-rater agreement was \textit{almost perfect} for both frameworks ($\kappa = .88$ for Epistemic, $\kappa = .91$ for ICAP), indicating a high level of consistency across coders.

\begin{table}
\centering
\caption{Epistemic Dimension and ICAP Engagement: Coding Scheme with Reliability}
\label{tab:codebook}
\begin{tabular}{p{1.8cm}p{9cm}p{1cm}p{1cm}}
\hline
\textbf{Code} & \textbf{Definition / Example} & \textbf{$\kappa$} & \textbf{$N$} \\
\hline
EP-PS & Problem Space: describes case facts or observations without invoking theoretical principles. Example: ``The nurses spent more time on tasks for Amy.'' & .87 & 554 \\
EP-CS & Conceptual Space: introduces or defines a general concept or principle. Example: ``Effective teamwork improves task management.'' & .70 & 554 \\
EP-CP-Adeq & Adequate Concept: correctly applies a principle to interpret the case. Example: ``Because they collaborated on Amy’s tasks, they managed stress more effectively.'' & .89 & 554 \\
EP-CP-Inad & Inadequate Concept: misapplies a concept or provides flawed reasoning. Example: ``They should have worked alone to finish faster.'' & .86 & 554 \\
EP-PK & Prior-Knowledge Relation: uses everyday or intuitive reasoning unrelated to domain theory. Example: ``Chocolate keeps you warm because sugar gives energy.'' & .71 & 554 \\
EP-OFF & Off-task: social or procedural statements not tied to the task. Example: ``ok,'' ``thanks.'' & .92 & 554 \\ \hline
\addlinespace[4pt]
I-PAS & Passive: minimal acknowledgment, no new content. Example: ``Okay.'' & .97 & 244 \\
I-ACT & Active: repeats or queries factual information. Example: ``What is the heart rate?'' & .89 & 244 \\
I-CON & Constructive: adds new inference or explanation beyond given content. Example: ``That suggests they coordinated to reduce stress.'' & .88 & 244 \\
I-INT & Interactive: builds upon or challenges AI reasoning collaboratively. Example: ``So if that’s true, the secondary nurse must have handled communication more.'' & --- & 244 \\
\hline
\end{tabular}
\end{table}

\subsection{Study Procedure}

\begin{figure}
    \centering
    \includegraphics[width=0.8\linewidth]{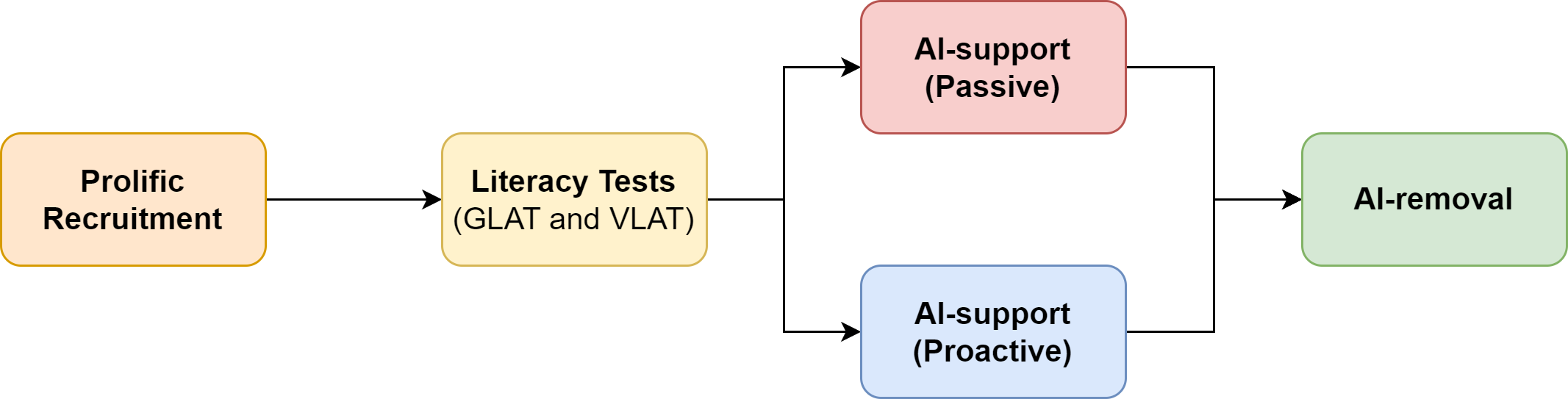}
    \caption{Experimental procedure consisting of two phases: (1) AI-supported writing task with either reactive or proactive agent assistance, and (2) AI-removal writing task without agent support.}
    \label{fig-design}
\end{figure}

We conducted a two-phase, mixed-method experimental study to examine how GenAI literacy and AI interaction type (reactive vs. proactive agent) influenced participants' academic writing performance with visual information. Participants were recruited via Prolific and directed to a Qualtrics\footnote{\url{https://www.qualtrics.com/}} survey, where they provided informed consent, demographics, and completed literacy assessments (\textit{GLAT} and \textit{mini-VLAT}). After reviewing contextual information on the healthcare simulation (Section~\ref{sec-materials}), they were randomly assigned to either the reactive or proactive agent assistance condition. In the AI-supported phase, participants composed a 150-word written response interpreting the provided multimodal visuals with real-time agent support. Subsequently, in the AI-removal phase, participants completed a similar task with new visual data from a different simulation scenario without agent access (Figure \ref{fig-design}). After completing both writing phases, participants answered a short post-task reflection questionnaire consisting of three Likert-type items and one open-ended question on their experiences with the agent (see Section~\ref{sec:rq4method}). The open-ended responses provided the qualitative dataset for the thematic analysis addressing RQ4. The study lasted approximately one hour. Ethical approval was obtained from [Anonymised] University (Project Number: Anonymised).

\subsection{Analytical Procedures}

\subsubsection{Epistemic network analysis (RQ1)}

\new{To examine how human--AI interaction design shaped the relational organisation of reasoning and engagement within dialogue, we applied epistemic network analysis (ENA; \citealt{shaffer_tutorial_2016}). ENA was selected because RQ1 concerned how coded ideas were connected within local conversational contexts, not only how frequently each code occurred. Frequency analysis would identify the prevalence of categories but not their patterned co-occurrence, while an immediate-transition approach would answer a different question about one-step sequence. ENA instead represents the structure of associations among discourse elements within a specified temporal window and permits comparison of participant-level networks between conditions.}

\new{We first modelled epistemic reasoning using the epistemic codes (EP-PS, EP-CS, EP-CP-Adeq, EP-CP-Inad, EP-PK, EP-OFF) across all human and AI utterances. A second model added ICAP engagement codes (I-PAS, I-ACT, I-CON, I-INT) for learner utterances to examine how engagement depth intersected with epistemic quality. Each utterance was represented as a binary vector indicating the presence or absence of each code. Each participant ($n = 79$) served as the unit of analysis, and co-occurrences were accumulated within a four-utterance moving window to capture local conversational context \citep{shaffer_tutorial_2016}. ENA generated a weighted adjacency matrix for each participant and projected the networks into two dimensions using mean rotation. MR1 maximised separation between the proactive and reactive conditions, and SVD2 captured orthogonal variance.}

\new{Nodes in ENA represent coded discourse categories, and an edge represents the strength with which two categories co-occurred within the moving window. A participant's point represents the centroid of that participant's network after projection; separation between condition centroids therefore indicates a difference in overall relational organisation rather than a difference in any single code frequency. Difference-network edges identify the associations that contributed most to condition separation. They should not be read as causal effects or as directional transitions from one code to another. Centroid positions were compared using Mann--Whitney $U$ tests with Holm adjustment, Cliff's delta quantified the magnitude of condition separation, and edge-weight contrasts were used to interpret the code-pair associations distinguishing the two interaction designs.}

\subsubsection{Regression analysis (RQ2)}

\new{To address RQ2, ordinal logistic regression analyses examined whether GenAI literacy predicted performance on the AI-removal task after accounting for domain-specific visualisation knowledge, agent condition, and performance on the corresponding AI-supported task. Separate models were fitted for \textit{Insightfulness}, \textit{Visual Data Integration}, \textit{Organisation and Coherence}, \textit{Linguistic Quality}, \textit{Critical Thinking}, and the composite outcome. The criterion ratings retained their ordered structure, justifying ordinal rather than ordinary least-squares regression. Predictors were standardised GenAI literacy ($\text{GenAI literacy}_i$), standardised domain knowledge ($\text{Domain knowledge}_i$; mini-VLAT), agent condition ($\text{Agent type}_i$; 0 = reactive, 1 = proactive), the participant's score on the same criterion during the AI-support task ($\text{AI-support score}_i$), and the interaction between GenAI literacy and agent condition. The model was specified as follows:}

\begin{equation}
\begin{split}
    \log\left(\frac{P(Y_i \leq k)}{1 - P(Y_i \leq k)}\right) = \beta_{0k} - \big(& \beta_{1}(\text{GenAI literacy}_i) \\
    & + \beta_{2}(\text{Domain knowledge}_i) \\
    & + \beta_{3}(\text{Agent type}_i) \\
    & + \beta_{4}(\text{AI-support score}_i) \\
    & + \beta_{5}(\text{GenAI literacy}_i \times \text{Agent type}_i) \big),
\end{split}
\end{equation}
\new{where $Y_i$ is the participant's ordered score on the AI-removal task and $k$ indexes the cumulative thresholds. GenAI literacy and visualisation literacy scores were standardised ($M = 0$, $SD = 1$), so the literacy odds ratios represent the change associated with a one-standard-deviation increase. Models used a logit link and maximum-likelihood estimation with the Broyden--Fletcher--Goldfarb--Shanno algorithm in the \texttt{Statsmodels} Python library. Diagnostics included variance inflation factors and the Brant test of the proportional-odds assumption. Statistical significance was set at $\alpha = .05$, and results were reported using coefficients, standard errors, $z$-values, $p$-values, and odds ratios. The literacy-by-condition term provided the formal test of whether the literacy slope differed between reactive and proactive designs.}

\subsubsection{Mediation analysis (RQ3)}

\new{To address RQ3, we examined whether the association between GenAI literacy and immediate independent writing performance was direct or operated indirectly through performance during the AI-supported task. GenAI literacy was measured before the writing tasks ($X$), AI-support performance occurred in the first writing phase ($M$), and AI-removal performance occurred in the second phase ($Y$), establishing the intended temporal ordering. The mediation analysis used two linear models. The first estimated the association between GenAI literacy and AI-support performance (path $a$), and the second estimated the associations of GenAI literacy (path $c'$) and AI-support performance (path $b$) with AI-removal performance. The indirect effect was $a \times b$:}

\begin{align}
M_i &= \alpha_M + a\,X_i + \varepsilon_{Mi}, 
\end{align}

\begin{align}
Y_i &= \alpha_Y + c'\,X_i + b\,M_i + \varepsilon_{Yi}.
\end{align}

\new{The indirect effect was estimated with 5{,}000 nonparametric bootstrap resamples and a 95\% confidence interval. Bootstrapping was used because the sampling distribution of a product term is often non-normal; an indirect effect was considered statistically distinguishable from zero only when its confidence interval excluded zero. The analysis was conducted separately for the reactive ($n = 39$) and proactive ($n = 40$) conditions to describe condition-specific patterns.}

\new{Interpretation of these models requires caution. The temporal sequence supports, but does not by itself establish, causal mediation because performance on the mediator was not randomised and the assumption of no unmeasured mediator--outcome confounding cannot be verified. The linear models also assume approximately linear additive relations and independent observations across participants. Bootstrap confidence intervals were the principal robustness procedure; no formal sensitivity analysis for unmeasured confounding was conducted. Accordingly, the mediation results are treated as exploratory, associational estimates, and differences between condition-specific coefficients are not considered evidence of moderation unless supported by the formal literacy-by-condition interaction test in the ordinal models.}

\subsubsection{Qualitative thematic analysis (RQ4)}
\label{sec:rq4method}

To address RQ4, we conducted a qualitative thematic analysis to derive evidence-based heuristics for human-AI interaction design from participants’ experiences with the reactive and proactive agents. After completing both AI-supported and independent writing tasks, participants answered an open-ended question: \textit{“Please describe your experience with the intervention. To what extent did it enhance or diminish your ability to interpret the data visualisations?”} This prompt aimed to elicit reflections on how the agent’s feedback, initiative, and interactional style influenced their analytical reasoning and sense of agency.

Following Braun and Clarke’s \citeyearpar{braun2006using} reflexive thematic analysis framework, the analysis proceeded in three iterative phases. First, all textual responses (N = 79) were read repeatedly to achieve familiarisation and initial inductive coding, with codes capturing recurring patterns related to feedback usefulness, cognitive engagement, and perceived autonomy. Second, codes were grouped and refined into higher-order themes that differentiated the learning experiences across the reactive and proactive designs. Finally, themes were collaboratively reviewed by the research team to ensure interpretive validity and coherence, following cross-verification procedures recommended by \citet{Guest_MacQueen_Namey_2012}. Throughout the analysis, a reflexive stance was maintained, and coding decisions were discussed until consensus was reached. The final themes were synthesised into three interaction design heuristics describing how reactive and proactive AI agents supported (or constrained) learner autonomy, reasoning, and equity in multimodal analytical writing.

\section{Results}\label{sec4}

\subsection{Relational Structure of Reasoning and Engagement}

\begin{figure}[!htbp]
    \centering
    \includegraphics[width=\linewidth]{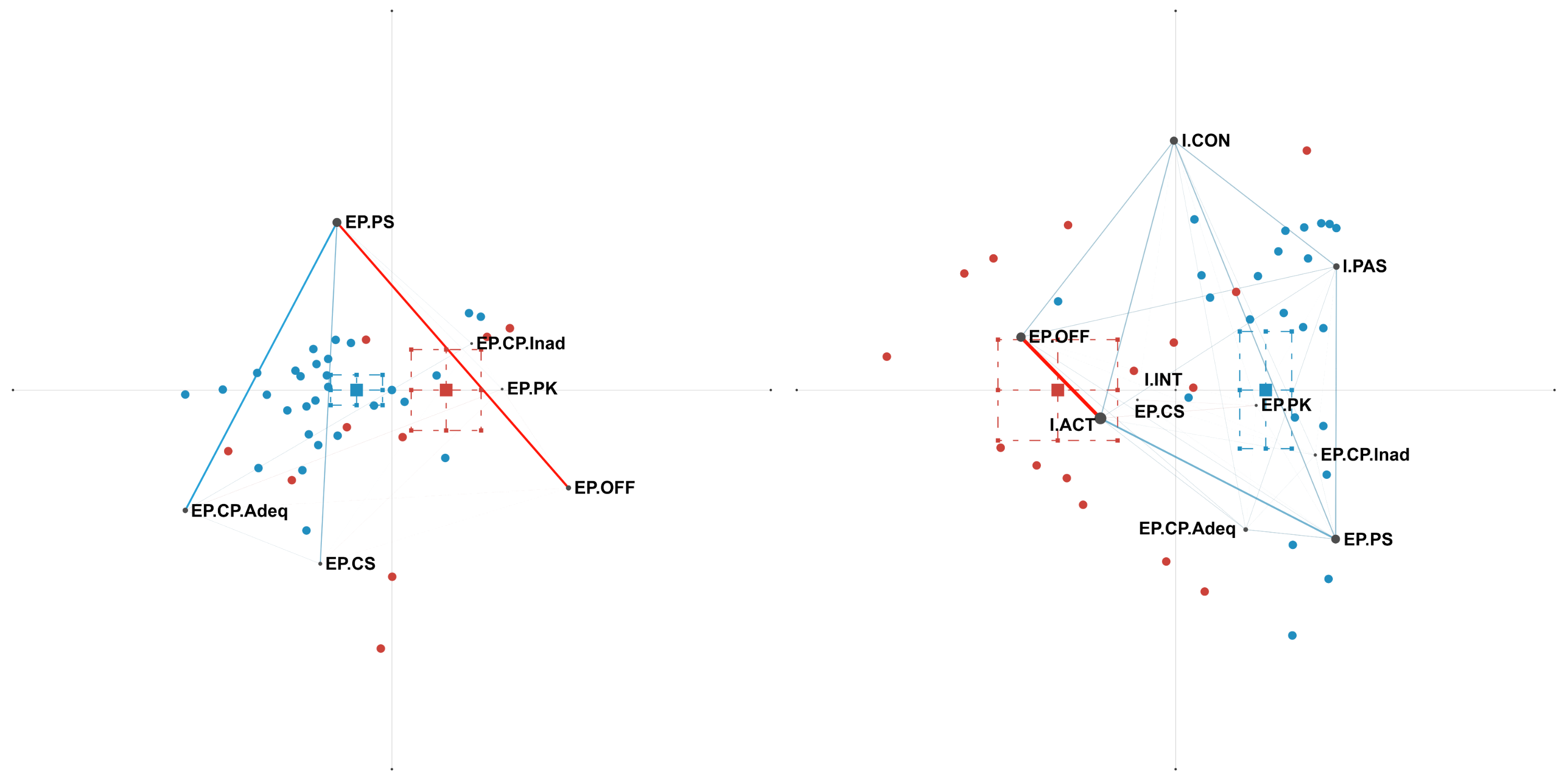}
    \caption{\new{Epistemic network analysis (ENA) results visualising the relational structure of reasoning and engagement in multimodal analytical writing. The left panel shows all speakers coded by epistemic dimensions; the right panel shows learner utterances combining epistemic and ICAP engagement codes. Blue edges indicate stronger co-occurrence associations in proactive interactions, and red edges indicate stronger associations in reactive interactions. Node size represents code frequency, and edge thickness represents co-occurrence strength within the moving window. Edges are relational associations rather than directional transitions or causal pathways.}}
    \label{fig:ena_epistemic_ic}
\end{figure}

\new{The ENA explained 38.3\% (MR1) and 22.7\% (SVD2) of variance in the discourse-network space. Reactive and proactive human--AI interaction designs were separated along MR1 ($W{=}228$, $p{<}.001$, Cliff's $\delta{=}{-}0.56$, large), indicating different configurations of reasoning and coordination (Figure~\ref{fig:ena_epistemic_ic}). Proactive agent dialogues exhibited stronger associations between conceptual and adequate reasoning (\textit{EP-CS}--\textit{EP-CP-Adeq}, $w{=}{-}0.12$) and between factual and adequate reasoning (\textit{EP-PS}--\textit{EP-CP-Adeq}, $w{=}{-}0.23$), reflecting more integrated epistemic elaboration. Reactive exchanges displayed a stronger factual/off-task association (\textit{EP-PS}--\textit{EP-OFF}, $w{=}0.27$), suggesting a more descriptive and procedural discourse organisation.}

\new{A secondary ENA including epistemic and engagement codes explained 27.5\% (MR1) and 26.4\% (SVD2) of variance. The conditions again differed along MR1 ($W{=}96$, $p{<}.001$, Cliff's $\delta{=}{-}0.78$, large). Proactive interactions showed stronger links between constructive engagement and adequate reasoning (\textit{I-CON}--\textit{EP-CP-Adeq}, $w{=}{-}0.13$) and between factual and constructive contributions (\textit{EP-PS}--\textit{I-CON}, $w{=}{-}0.13$). Reactive dialogues emphasised procedural pairings such as \textit{EP-OFF}--\textit{I-ACT} ($w{=}0.43$) and \textit{EP-PS}--\textit{I-ACT} ($w{=}{-}0.22$).}

\new{Educationally, the large centroid separations indicate that the two designs changed how ideas were connected during interaction, not simply how often a category appeared and not necessarily the final score achieved. The proactive sequence appears to have externalised part of the planning and monitoring work by repeatedly directing attention from visual features to interpretations and cross-visual connections. The reactive design left this orchestration to the learner, producing a greater emphasis on requesting or confirming factual information. Learner reflections provide a convergent process explanation: one proactive-condition participant reported that the agent ``asked the right questions on what to focus on'' (P18), while another said that it helped relate the visuals to one another (P17). These accounts help interpret the network patterns but are not treated as causal confirmation.}

\subsection{GenAI Literacy and Writing Outcomes}

\new{All ordinal logistic regression models met the proportional-odds assumption (Brant tests, all $p > .05$) and showed no problematic multicollinearity (all variance inflation factors $< 5$). GenAI literacy significantly predicted three dimensions of immediate independent performance after AI support was removed: \textit{Visual Data Integration}, $b = 0.13$, $SE = 0.06$, $z = 2.07$, $p = .039$, $OR = 1.14$; \textit{Critical Thinking}, $b = 0.14$, $SE = 0.07$, $z = 2.18$, $p = .029$, $OR = 1.15$; and the overall \textit{Composite Score}, $b = 0.10$, $SE = 0.05$, $z = 2.15$, $p = .032$, $OR = 1.11$. GenAI literacy did not significantly predict \textit{Insightfulness}, \textit{Organisation and Coherence}, or \textit{Linguistic Quality}.}

\new{Because GenAI literacy was standardised, these odds ratios correspond to a one-standard-deviation increase in literacy. The estimated associations were modest: the odds of a higher score increased by 11--15\%, with the clearest relationships in integrating visual evidence and evaluating implications. Thus, literacy was more closely related to higher-order evidence use than to language form or organisation. These are educationally relevant but small-to-moderate shifts in immediate task performance, not evidence of long-term learning.}

\new{Performance during the AI-supported task was a consistent predictor of performance during the subsequent AI-removal task. Higher AI-support scores predicted independent performance for \textit{Insightfulness}, $b = 1.67$, $SE = 0.53$, $z = 3.18$, $p = .001$; \textit{Visual Data Integration}, $b = 2.61$, $SE = 0.59$, $z = 4.39$, $p < .001$; \textit{Organisation and Coherence}, $b = 1.52$, $SE = 0.49$, $z = 3.11$, $p = .002$; \textit{Linguistic Quality}, $b = 2.65$, $SE = 0.57$, $z = 4.63$, $p < .001$; \textit{Critical Thinking}, $b = 3.57$, $SE = 0.77$, $z = 4.67$, $p < .001$; and the \textit{Composite Score}, $b = 0.61$, $SE = 0.12$, $z = 5.03$, $p < .001$. This continuity indicates that students who performed well with support also tended to perform well immediately afterwards, although it cannot distinguish learning from stable differences in writing competence.}

\new{Neither domain-specific knowledge nor agent condition had a significant unique association with AI-removal performance, and none of the GenAI-literacy-by-agent interactions was significant. The formal models therefore did not support different literacy slopes in the reactive and proactive conditions. This null interaction is important for interpretation: condition-specific coefficient patterns reported below are descriptive and should not be treated as evidence that agent design statistically moderated the literacy effect. Because an interaction visualisation could imply a separation not supported by the inferential test, we report the interaction result directly rather than emphasising simple-slope plots.}

\subsection{Mediation Effects of GenAI Literacy}

\new{In the reactive condition, GenAI literacy had a significant total association with AI-removal performance (path $c$: $\beta = 0.172$, $p = .020$; Table~\ref{tab:mediation}). The indirect effect through AI-support performance was small and its 95\% confidence interval included zero (indirect effect $= 0.032$, 95\% CI $[-0.022, 0.108]$). The direct coefficient remained positive after the mediator was included (path $c'$: $\beta = 0.140$), so the observed literacy association was not explained by the indirect pathway.}

\new{In the proactive condition, neither the total coefficient (path $c$: $\beta = 0.029$) nor the direct coefficient (path $c'$: $\beta = 0.007$) was significant. The indirect effect was also small and uncertain (indirect effect $= 0.022$, 95\% CI $[-0.055, 0.102]$). Descriptively, this pattern is consistent with smaller literacy-related performance differences in the proactive condition. However, both indirect-effect intervals included zero and the formal literacy-by-condition interactions in the ordinal models were non-significant. The results therefore do not establish mediation, compensation, or moderation by proactive support.}

\new{The educational interpretation is correspondingly cautious. Reactive interaction required learners to decide what help to request and how to evaluate it, so a direct association with GenAI literacy is plausible. Proactive participants described receiving targeted questions and process guidance, which offers a possible explanation for the flatter condition-specific pattern. Yet the sample was modest, the confidence intervals were wide, and learner reflections were not analysed as measured literacy subgroups. The condition-specific estimates should therefore be viewed as hypothesis-generating evidence for future, adequately powered tests of adaptive initiative.}

\begin{table}[h!]
\centering
\caption{Bootstrap mediation analysis results by condition}
\label{tab:mediation}
\begin{tabular}{lrr}
\hline
\textbf{Parameter} & \textbf{Reactive} & \textbf{Proactive} \\
\hline
\textit{Paths} & & \\
\quad Path a (X $\rightarrow$ M) & 0.069 & 0.051 \\
\quad Path b (M $\rightarrow$ Y | X) & 0.468 & 0.438 \\
\quad Path c' (Direct effect) & 0.140 & 0.007 \\
\quad Path c (Total effect) & 0.172* & 0.029 \\
\hline
\textit{Indirect effect (a $\times$ b)} & & \\
\quad Point estimate & 0.032 & 0.022 \\
\quad 95\% confidence interval & [-0.022, 0.108] & [-0.055, 0.102] \\
\quad Significant & No & No \\
\hline
\multicolumn{3}{l}{\textit{Note.} *\(p < .05\). X = GenAI literacy, M = AI-support score, Y = AI-removal score.}
\end{tabular}
\end{table}

\subsection{Design Heuristics for Agent Interactivity}

\new{Thematic analysis of participants' reflections was used to explain how the interaction patterns identified quantitatively were experienced by learners, rather than to claim independent causal confirmation. The accounts revealed two distinct yet complementary interaction patterns that shaped reasoning, feedback use, and perceived autonomy. Three empirically grounded heuristics emerged: (1) sustain autonomy through contextual and confirmatory feedback, (2) promote integrative reasoning and immediate independent application through dialogic scaffolding, and (3) align initiative with perceived learner expertise and task complexity.}

\paragraph{Heuristic 1: Sustain autonomy through contextual and confirmatory feedback (Reactive).}
Participants described the reactive agent as a responsive but non-intrusive partner that provided context and reassurance without taking over the interpretive process. Ten participants emphasised how it clarified relationships between visual features and the underlying scenario:  
\textit{“The intervention provided valuable context for interpreting data visualisations by highlighting key aspects of task prioritisation, communication, and stress levels among nurses. It effectively guided my understanding of how these factors influenced performance”} (P31).  
Others appreciated how it translated visual patterns into concise textual explanations, aiding comprehension:  
\textit{“I felt that the intervention helped clarify some aspects of the data visualisations. For instance, I was able to learn the significance of the thickness of the lines in the communication network”} (P98).  
Several participants also valued the agent’s confirmatory role, noting that it validated their interpretations and made complex visuals easier to articulate:  
\textit{“It greatly enhanced my understanding by confirming what I was already seeing and converting the images to words for easy and better comprehension”} (P108).  
Efficiency and convenience were recurring benefits:  
\textit{“I was surprised by how efficiently it addressed my queries. It made my work easier and saved me a lot of time”} (P72).  
However, participants with higher visualisation literacy found the assistance redundant, remarking that \textit{“It did not contribute much as it provided no additional insight; using the agent was slower and less straightforward than analysing the graphs”} (P21).  
These reflections suggest that reactive feedback supports autonomy and confidence by confirming learners’ reasoning, but its usefulness decreases when tasks are simple or learners are already proficient.

\paragraph{\new{Heuristic 2: Promote integrative reasoning and immediate independent application through dialogic scaffolding (Proactive).}}
Learners working with the proactive agent described a more interactive and cognitively demanding experience in which the AI initiated questions, offered structured guidance, and prompted connections across multiple visual sources. Many participants (N = 14) noted that the agent helped them synthesise information into holistic interpretations:  
\textit{“I am able to put together different components that are given and make one conclusion rather than look at one part and just make a conclusion”} (P48);  
\textit{“It allowed me to relate the visuals to one another in order to maximise my understanding of the situation”} (P17).  
Ten participants highlighted the value of targeted guidance and process-level feedback:  
\textit{“The intervention enhanced my ability to interpret the data visualisation as it provided guidance and understanding of the time allocation, stress level and communication pattern”} (P101);  
\textit{“It provided meaningful feedback and asked the right questions on what to focus on and what you have to get right”} (P18).  
Others emphasised how the agent fostered reflection and self-correction:  
\textit{“It was helpful to consolidate things our brain may have already been assuming, and it told us when we were wrong about these diagrams or images”} (P27).  
Nevertheless, a few participants felt that the proactive feedback was excessive for simple tasks:  
\textit{“In the first question, it only confirmed what could be observed easily in the chart”} (P8),  
or unnecessary given their competence:  
\textit{“I thought it was helpful in guiding me along. However, I felt as though it was not completely necessary. I would value it if I needed help, but I don’t know if I really did”} (P56).  
These accounts indicate that proactive interaction deepened conceptual integration and metacognitive reflection, but the degree of guidance should scale with task complexity to prevent over-scaffolding.

\paragraph{Heuristic 3: Ensure equity through adaptive alignment with learner expertise and task complexity.}
\new{Comparing experiences across both designs revealed differences in perceived benefit associated with participants' self-described confidence and task difficulty. Participants who wanted more support credited the proactive agent for step-by-step guidance (\textit{``It asked the right questions on what to focus on and what you have to get right''}; P18), whereas a participant who regarded the visual task as straightforward preferred greater autonomy (\textit{``Using the agent was slower and less straightforward than analysing the graphs''}; P21). These reflections are consistent with the alignment account, but they do not establish that the comments map directly onto measured GLAT scores.}

\new{The divergence identifies a practical equity risk: a uniform interaction style may under-support some learners while over-directing others. A design implication for future AI agents is to allow initiative to vary using transparent cues such as an explicit request for more guidance, repeated clarification, or demonstrated completion of prior steps. Such adaptation should preserve the learner's ability to skip, question, or reduce scaffolding rather than inferring competence opaquely.}

\section{Discussion}\label{sec5}

\new{This study examined how a reactive versus proactive distribution of agent initiative related to multimodal analytical writing and how GenAI literacy was associated with immediate performance after support was removed. The evidence comes from one short, controlled writing context, so the discussion treats the agency gap as an emerging explanatory framework rather than an established construct. Across methods, the clearest result was a distinction between process and outcome: agent design produced large differences in the relational organisation of dialogue, whereas its direct association with immediate writing scores and its interaction with literacy were not statistically significant.}

\new{For RQ1, proactive dialogues showed stronger connections among conceptual statements, adequate application, and constructive engagement. This pattern is consistent with dialogic scaffolding that directs attention from observations to principles and synthesis \citep{weinberger_framework_2006, bai_2024, Looi_2025}. Reactive dialogue preserved learner control over when and what to ask, but its network was organised more around factual and procedural exchanges. The educational meaning is not that proactive dialogue was universally ``deeper'' or that reactive dialogue was deficient. Rather, proactive design externalised some planning and monitoring functions, while reactive design made the learner's own regulation more visible. This supports treating initiative allocation as an interaction-design variable that shapes opportunities for epistemic elaboration, beyond model accuracy alone.}

\new{For RQ2, GenAI literacy had modest associations with visual data integration, critical thinking, and the composite score on the immediate AI-removal task. These dimensions require learners to evaluate, connect, and use evidence, which aligns with the evaluative and procedural capabilities represented by the GLAT \citep{jin2024glat, Ng_2021}. The effects did not extend to every writing dimension, and the task occurred directly after the supported phase. The result should therefore be described as immediate independent application or near transfer, not durable learning transfer. The non-significant condition and interaction terms further indicate that the study did not demonstrate that one design produced higher scores or different literacy slopes.}

\new{For RQ3, the condition-specific mediation estimates are suggestive but not confirmatory. A total literacy association appeared in the reactive condition and not in the proactive condition, which is consistent with the idea that a highly learner-initiated interface may place greater demands on prompting and evaluation skills. However, neither indirect effect was significant, and the formal literacy-by-condition tests were non-significant. The findings therefore do not show that proactive scaffolding compensated for low literacy or that design moderated a causal pathway. They instead motivate a future hypothesis: when an interface supplies more of the questioning sequence, observable performance differences associated with literacy may become smaller during an immediate task. Testing that hypothesis requires larger samples, direct measures of learner agency, and experimentally varied levels of initiative.}

\new{For RQ4, participant reflections help explain why initiative may need calibration. Learners valued reactive feedback when it confirmed an interpretation without interrupting their process, and they valued proactive feedback when it connected evidence across visuals and prompted self-correction. They also described proactive guidance as redundant when the task appeared easy. These experiences resemble adaptive-scaffolding and expertise-reversal accounts \citep{van2010scaffolding, kalyuga2009expertise}, but the agency-gap framing adds a specific human--AI design question: whether the system's share of initiation leaves the learner with a productive amount of epistemic responsibility.}

\subsection{Implications for Educational Research}

\new{The principal theoretical contribution is a relational account that links AI agent initiative and learner GenAI literacy to the distribution of epistemic work. This account is distinct from learner agency, which names the learner's capacity to act; self-regulated learning, which describes regulatory processes; shared regulation, which describes how those processes are distributed among partners; and adaptive scaffolding, which is a mechanism for adjusting instructional support \citep{Panadero_2017, Hadwin_2018, van2010scaffolding}. The agency gap focuses more narrowly on a mismatch between the initiative embedded in an AI interface and the capability required to use that interface while retaining meaningful control. The present findings provide preliminary observable implications of that account, not a validation of a new scale or general theory. Future studies should test this framework directly. Designs could factorially manipulate agent initiative and scaffold adaptivity, stratify or continuously model GenAI literacy, and include independent measures of learner agency, self-regulation, cognitive load, and feedback uptake. Longitudinal work should examine whether scaffolds are internalised, whether initiative can be faded without performance loss, and whether literacy itself changes through repeated interaction. Confirmatory research should also preregister the hypothesised interaction and mediation models and recruit samples large enough to estimate small indirect and interaction effects with precision. The underlying mechanism may be relevant beyond medical and nursing education wherever writing requires evidence selection and synthesis, such as business analytics, social-science reports, laboratory explanations, engineering design rationales, or data journalism. Generalisation cannot be assumed, however, because disciplines differ in what counts as evidence, how uncertainty is expressed, and how visuals should be integrated into argument. Cross-disciplinary replication should therefore preserve the initiative manipulation while adapting the scaffold content and assessment rubric to local epistemic norms.}

\subsection{Implications for Educational Practice}

\new{For writing tasks similar to those studied here, educators and designers can translate the findings into three cautious design practices. First, make the level of initiative visible and adjustable: learners should be able to request a question sequence, pause it, skip a prompt, or return to on-demand help. Second, structure proactive support around a transparent progression, orient to the representation, identify evidence, interpret relationships, connect sources, and synthesise, rather than asking the model to provide an answer. Third, fade or reduce prompts when learners demonstrate that they can identify relevant evidence and justify an interpretation independently. These practices preserve opportunities for learner control while offering support when the task or interface would otherwise impose excessive initiation demands. GenAI literacy should also be taught as part of academic writing rather than assumed. Short activities can require students to formulate and compare prompts, identify unsupported AI claims, justify which AI suggestions they accepted or rejected, and revise a response without AI after a supported draft. For AI agent procurement and implementation, institutions should require transparency about when proactive interventions are triggered, what learner data are used, and how learners can override automated support. The present study does not justify deploying fully adaptive learner profiling, but it supports giving users explicit control over the depth and frequency of scaffolding.}

\subsection{Limitations and Future Directions}

\new{Several limitations constrain the strength and scope of the conclusions. First, the sample of 79 was planned for the primary two-condition comparison, not for every secondary regression and mediation parameter. The ordinal models were parsimonious and passed reported diagnostics, but estimates of modest associations may be unstable. The condition-specific mediation analyses had only 39--40 participants, and their wide bootstrap confidence intervals showed limited precision for small indirect effects. Those analyses should be considered exploratory until replicated in larger samples. Second, participants were medical or nursing students and recent graduates recruited through an online platform. Their familiarity with healthcare scenarios, disciplinary conventions, and GenAI may differ from learners in other fields or classroom settings. The results therefore generalise most directly to short multimodal evidence-synthesis tasks with similar participants, not to academic writing as a whole. Third, the AI-removal task followed immediately after the supported task. It measures short-term independent performance on a related task, but it cannot establish retention, delayed transfer, or the development of enduring writing competence. Fourth, the agency gap was theorised rather than directly measured. Learner agency was inferred from interaction patterns and reflections, and no standalone agency or self-regulation scale was administered. Similarly, although the prompt specified different initiative rules, the study did not report an independent manipulation-check coding of the realised agent turns. Future research should measure perceived and behavioural agency and audit intervention fidelity, including the frequency, timing, and type of proactive moves. Fifth, the GLAT analysis relied on published validation evidence; future studies should report sample-specific reliability and examine whether the measure functions equivalently across disciplines and experience levels. Finally, natural-language generation can vary in wording even under a fixed prompt, and mediation models cannot rule out unmeasured differences that influence both supported and independent performance. Future work should combine larger preregistered experiments with process logs, delayed assessments, and sensitivity analyses. It should also compare fixed reactive and proactive designs with transparent adaptive strategies that use explicit learner choices or observable task progress rather than opaque competence inference.}

\section{Conclusion}\label{sec6}

\new{This study contributes a relational perspective on human--AI interaction in writing: the educational consequences of GenAI depend not only on what a model can generate, but also on how conversational initiative is divided between learner and system. In this short multimodal writing context, proactive design reorganised dialogue towards stronger conceptual, evidential, and constructive connections, while GenAI literacy showed modest associations with selected dimensions of immediate independent performance. The direct condition effect and literacy-by-condition interactions were not significant, and the condition-specific mediation patterns were exploratory. The \textit{agency gap} therefore remains an emerging conceptualisation: a mismatch in which an interface either demands more self-initiation than a learner can currently exercise or supplies so much direction that opportunities for learner control are reduced. Its theoretical value is to connect agent initiative, learner capability, and the allocation of epistemic work in a testable framework. For educators and designers, the practical implication is to make initiative adjustable, use proactive prompts to orient, interpret, connect, and synthesise rather than to supply answers, and reduce scaffolding as learners demonstrate independent control. Future research should test these principles across disciplines and learner populations, include direct measures of agency and regulation, and examine delayed learning and transfer under adaptive and fading support.}

\appendix
\section{Appendix}
\section{Generative AI Literacy Assessment Test (GLAT)}
\label{sec:appendix-GLAT}
\begin{footnotesize}
\renewcommand{\arraystretch}{1.2}
\begin{longtable}{|p{0.5cm}|p{1.5cm}|p{5cm}|p{7cm}|}
\hline
\textbf{Item} & \textbf{Dimension} & \textbf{Question} & \textbf{Options (Answer Bolded)}\\
\hline
1 & Know \& Understand & Which of the following best describes "Generative AI"? & 
\textbf{A. AI that creates new content like text, images, or music by learning from existing data.} \newline
B. An AI system designed to enhance the speed and accuracy of data retrieval in search engines. \newline
C. A form of artificial intelligence that focuses on translating languages in real-time. \newline
D. AI technology used primarily for managing and organizing large databases. \\
\hline

2 & Know \& Understand & Which of the following statements best describes an LLM (Large Language Model)? & 
A. It generates text by analyzing and summarizing large volumes of web content. \newline
\textbf{B. It generates text by predicting the next word based on the context of previous words.} \newline
C. It generates text by translating input text into multiple languages simultaneously. \newline
D. It generates text by using pre-defined templates and filling in the blanks. \\
\hline

3 & Know \& Understand & Which of the following tasks can Generative AI perform with a high degree of accuracy? & 
A. Predicting stock market trends \newline
B. Making ethical decisions in complex scenarios \newline
C. Diagnosing rare diseases \newline
\textbf{D. Generating human-like text based on prompts} \\
\hline

4 & Know \& Understand & In the context of Generative AI, what is "zero-shot learning"? & 
A. Training a model without any data. \newline
\textbf{B. The ability of a model to perform a task without any task-specific training.} \newline
C. A method of reducing the model's training time to zero. \newline
D. A technique for generating synthetic training data. \\
\hline

5 & Know \& Understand & Which of the following is a potential challenge when using prompt-based development for text generation? & 
A. The language model can only generate binary outputs. \newline
B. The need for extensive labeled data to train the model. \newline
\textbf{C. Crafting a prompt that accurately captures the desired context and nuances.} \newline
D. The requirement for complex feature engineering. \\
\hline

6 & Know \& Understand & What does the term "token" refer to in the context of a large language model (LLM)? & 
\textbf{A. A token is a unit of text, such as a word or a subword, that the model processes individually.} \newline
B. A token is a unique identifier assigned to each user interacting with the language model. \newline
C. A token is a security measure used to authenticate API requests to the language model. \newline
D. A token is a reward given to users for contributing valuable data to train the language model. \\
\hline

7 & Know \& Understand & Which of the following is NOT a requirement for an AI to be considered artificial general intelligence (AGI)? & 
A. The ability to learn and adapt to new tasks without human intervention. \newline
B. The capability to perform tasks across various domains with human-like proficiency. \newline
\textbf{C. The ability to predict future events with perfect accuracy.} \newline
D. The capacity to understand and generate natural language. \\
\hline

8 & Know \& Understand & How does RAG (Retrieval-Augmented Generation) enhance the capabilities of an LLM? & 
A. By improving its grammar and syntax. \newline
\textbf{B. By providing it with real-time and relevant data.} \newline
C. By increasing its computational speed. \newline
D. By enabling it to understand multiple languages. \\
\hline

9 & Use \& Apply & When using generative AI to create a marketing pitch, which of the following strategies is least likely to be effective? & 
A. Supplying the AI with information about the target audience \newline
B. Asking the AI to include unique selling points and benefits \newline
C. Requesting the AI to use persuasive language techniques \newline
\textbf{D. Providing the AI with a list of competitors' products} \\
\hline

10 & Use \& Apply & After deploying a customer service agent, you notice that it frequently provides outdated information about company policies. What is the best course of action to address this issue? & 
A. Implement a feedback loop where users can flag outdated information for review. \newline
\textbf{B. Schedule regular updates to the agent's training data to include the latest company policies.} \newline
C. Set up a system where complex or policy-related queries are escalated to human agents for accurate responses. \newline
D. Conduct a comprehensive audit of the agent's performance metrics to identify areas for improvement. \\
\hline

11 & Use \& Apply & Suppose you have a large dataset of emails and you want to build an application to answer questions based on this dataset. Which of the following scenarios best illustrates the advantage of using RAG over prompting (i.e., without RAG)? & 
A. You need to generate creative writing pieces based on the email content. \newline
B. You want to ensure the model can answer questions even if it has never seen similar questions before. \newline
\textbf{C. You need to answer questions that require specific information from different parts of the email dataset.} \newline
D. You want to reduce the size of the language model to save computational resources. \\
\hline

12 & Evaluate \& Create & As a student using a Large Language Model (LLM) to gather information for an assignment, how should you approach the information it provides? & 
A. The LLM's answers are always more trustworthy than any information you will find on the internet, so you can use them without further verification. \newline
B. The LLM's answers are generally more trustworthy than internet sources, but you should still verify the information with other reliable sources. \newline
\textbf{C. The LLM's answers are not necessarily more trustworthy than internet sources, and you should cross-check the information with other credible references.} \newline
D. The LLM's answers are less trustworthy than internet sources because it relies on outdated information. \\
\hline

13 & Evaluate \& Create & It is unlikely for an LLM to provide an accurate summary of the latest financial market trends in real-time. Is this statement true or false? & 
\textbf{A. True, because the LLM's data may be outdated due to its knowledge cutoff.} \newline
B. True, because the LLM is not good at handling numbers and structured data. \newline
C. False, because the LLM frequently updates its knowledge base. \newline
D. False, because the LLM is capable of synthesizing the latest market data automatically. \\
\hline

14 & Evaluate \& Create & A generative AI tool has provided a summary of a research paper. The summary states, "The study found that increased screen time is directly correlated with decreased attention spans in children aged 8-12." What is your next step? & 
A. Accept the summary as accurate because AI tools are generally reliable. \newline
B. Ask the AI to provide more details about the study's methodology and results. \newline
\textbf{C. Cross-check the summary with the original research paper.} \newline
D. Use another AI tool to generate a summary for comparison and evaluate the consistency between both summaries \\
\hline

15 & Evaluate \& Create & While reviewing a video of a well-known public figure making controversial statements, which characteristic confirms the video was NOT generated by AI? & 
A. The public figure's voice sounds like themselves. \newline
B. The video has a professional and polished appearance. \newline
C. The video is high-quality with smooth transitions. \newline
\textbf{D. None of the above.} \\
\hline

16 & Ethics & When a generative AI system is used for screening job applications, what issue might arise concerning the quality and fairness of hiring decisions? & 
A. The AI system might overlook applicants' unique achievements and extracurricular activities. \newline
B. The AI system could misinterpret minor formatting differences in resumes. \newline
C. The AI system might not effectively handle applications submitted in various languages. \newline
\textbf{D. The AI system could reinforce existing biases found in historical hiring data.} \\
\hline

17 & Ethics & In a healthcare startup, an accurate AI model recommends treatments, but doctors don't trust it because they can't understand how the model arrived at its conclusions. What core issue does this scenario illustrate? & 
A. The AI model uses obsolete training data. \newline
B. The training dataset lacks sufficient diversity. \newline
C. The treatment guidelines input are incorrect. \newline
\textbf{D. The AI model behaves as a black box.} \\
\hline

18 & Ethics & What are the potential copyright implications for a journalist using an AI-generated image in a commercial article? & 
\textbf{A. The journalist needs to check the licensing policy of the AI tool they used.} \newline
B. The AI-generated image is automatically free to use without any restrictions. \newline
C. The journalist must pay a standard licensing fee to use the AI-generated image. \newline
D. The image cannot be used in any commercial context because it is AI-generated. \\
\hline

19 & Ethics & Should we impose restrictions on the outputs of generative AI technologies? & 
A. Yes, to reduce the computational resources required for operating these technologies. \newline
\textbf{B. Yes, to prevent the dissemination of harmful or misleading content.} \newline
C. No, as it would hinder technological innovation and creativity. \newline
D. No, because users should have the freedom to access all generated content. \\
\hline

20 & Ethics & Sending personal information to cloud-based generative AI tools has little privacy concerns. & 
A. True, as this information is encrypted using sophisticated algorithms during the transmission process. \newline
B. True, as generative AI tools are black-box systems and cannot output personal information even if it is used for model training. \newline
\textbf{C. False, as generative AI tools train on unencrypted data and can output private information based on their probabilistic nature.} \newline
D. False, as advancements in quantum computing can easily decipher the encrypted data. \\
\hline
\end{longtable}
\end{footnotesize}

\section{Mini-Visualization Literacy Assessment Test (VLAT)}
\label{sec:appendix-vlat}
\begin{footnotesize}
\renewcommand{\arraystretch}{1.2}
\begin{longtable}{|p{2cm}|p{3cm}|p{10.5cm}|}
\hline
\textbf{Item ID} & \textbf{Visualization Type} & \textbf{Question} \\

\hline
1 & Line Chart & What was the price of a barrel of oil in February 2020? \\
\hline
2 & Bar Chart & What is the average internet speed in Japan? \\
\hline
3 & Stacked Bar Chart & What is the cost of peanuts in Seoul? \\
\hline
4 & 100\% Stacked Bar Chart & Which country has the lowest proportion of Gold medals? \\
\hline
5 & Pie Chart & What is the approximate global smartphone market share of Samsung? \\
\hline
6 & Histogram & What distance have customers traveled in the taxi the most? \\
\hline
7 & Scatterplot & There is a negative linear relationship between the height and the weight of the 85 males. \\
\hline
8 & Area Chart & What was the average price of pound of coffee beans in October 2019? \\
\hline
9 & Stacked Area Chart & What was the ratio of girls named ``Isla'' to girls named ``Amelia'' in 2012 in the UK? \\
\hline
10 & Bubble Chart & Which city’s metro system has the largest number of stations? \\
\hline
11 & Choropleth Map & In 2020, the unemployment rate for Washington (WA) was higher than that of Wisconsin (WI). \\
\hline
12 & Treemap & eBay is nested in the Software category. \\
\hline

\end{longtable}
\end{footnotesize}

\section{Rubric for Evaluating Academic Writing with Visual Data}
\label{sec:appendix-rubric}
\begin{footnotesize}
\begin{longtable}{@{}L{3cm}L{3cm}L{10cm}@{}}
\toprule
\textbf{Metric} & \textbf{Score} & \textbf{Description and Example} \\
\midrule
\endfirsthead

\toprule
\textbf{Metric} & \textbf{Score} & \textbf{Description and Example} \\
\midrule
\endhead

\midrule
\multicolumn{3}{r}{\footnotesize Continued on next page} \\
\endfoot

\bottomrule
\endlastfoot

\multirow{5}{*}{Insightfulness}
& Level 5 (Excellent) &
- Provides highly original and profound insights.\newline
- Provides a holistic response that addresses every key element of the analysis.\newline
- Draws thought-provoking, actionable conclusions that extend beyond obvious interpretations.\newline
\textbf{Example:} \emph{"Nurse 1's disproportionate focus on Amy (50\%) coupled with their central communication role and elevated stress levels (111 BPM near Bed 4) indicates an unsustainable workload distribution. Implementing a more balanced task allocation could enhance team efficiency and reduce individual stress."} \\
\cmidrule(lr){2-3}
& Level 4 (Very Good) &
- Offers detailed and valid analysis with some synthesis of data.\newline
- Provides a well-rounded response but may lack depth in one or two areas or overlook minor elements.\newline
- Conclusions are solid but may lack the depth or originality of Level 5.\newline
\textbf{Example:} \emph{"While focusing on Amy aligns with priorities, Nurse 1's increased stress due to communication demands suggests that sharing responsibilities might alleviate stress and improve outcomes."} \\
\cmidrule(lr){2-3}
& Level 3 (Good) &
- Provides accurate analysis based on direct observations.\newline
- Limited synthesis of data; basic, isolated analysis with little connections.\newline
- Insights are correct but straightforward and may be somewhat obvious.\newline
\textbf{Example:} \emph{"Nurse 1 spent more time attending to Amy, which led to higher stress levels compared to Nurse 2."} \\
\cmidrule(lr){2-3}
& Level 2 (Fair) &
- Makes surface-level observations with minimal analysis.\newline
- Omits key elements or provides superficial explanations.\newline
- Demonstrates a limited understanding of the scenario.\newline
\textbf{Example:} \emph{"Nurse 1 worked on Amy for half the time and had a higher heart rate."} \\
\cmidrule(lr){2-3}
& Level 1 (Poor) &
- Provides an incomplete or highly superficial response.\newline
- Lacks any meaningful interpretation or insight.\newline
\textbf{Example:} \emph{"Nurse 1 was stressed with a higher heart rate while spending time on Amy."} \\
\midrule

\multirow{5}{*}{Visual Data Integration}
& Level 5 (Excellent) &
- Fully integrates all relevant visual data types into the analysis.\newline
- Draws meaningful connections between different data sources.\newline
- Builds a cohesive narrative that supports complex conclusions.\newline
\textbf{Example:} \emph{"The solo focus (50\%) on Amy, Nurse 1's central role in communication, and elevated heart rate (111 BPM at Bed 4) collectively highlight an imbalance in task distribution, emphasising the need for better teamwork."} \\
\cmidrule(lr){2-3}
& Level 4 (Very Good) &
- Effectively incorporates multiple types of visual data.\newline
- Makes connections between data types, though some integrations may be less thorough.\newline
- Uses visual data to support analysis, but may not fully exploit all available information.\newline
\textbf{Example:} \emph{"Nurse 1's extensive time with Amy and higher stress levels near Bed 4 suggest that task allocation affected their stress, as shown in the charts and heart rate data."} \\
\cmidrule(lr){2-3}
& Level 3 (Good) &
- Utilises at least one type of visual data effectively.\newline
- References other data types but with limited integration.\newline
- Analysis supported by visual data but lacks depth in synthesis.\newline
\textbf{Example:} \emph{"According to the communication network, Nurse 1 talked more, which might explain their higher stress level."} \\
\cmidrule(lr){2-3}
& Level 2 (Fair) &
- Mentions visual data superficially without meaningful integration.\newline
- Relies on single data points without connecting them to the broader context.\newline
\textbf{Example:} \emph{"The chart shows Nurse 1 had a heart rate of 111 BPM near Amy."} \\
\cmidrule(lr){2-3}
& Level 1 (Poor) &
- Fails to incorporate visual data into the analysis.\newline
- May misinterpret or ignore available data.\newline
\textbf{Example:} \emph{"Nurse 1 was stressed, as seen in the data."} \\
\midrule

\multirow{5}{*}{Organisation and Coherence}
& Level 5 (Excellent) &
- Presents ideas in a clear, logical order with excellent structure.\newline
- Uses seamless transitions between points and paragraphs.\newline
- Arguments are well-developed and interconnected.\newline
\textbf{Example:} \emph{"By allocating her solo efforts (50\%) to Amy, Nurse 1 became central in patient care. This, combined with her elevated stress levels (111 BPM near Bed 4) and primary communication role, indicates a need for task redistribution to promote balance and reduce burnout."} \\
\cmidrule(lr){2-3}
& Level 4 (Very Good) &
- Maintains a clear and logical structure with minor lapses.\newline
- Uses transitions effectively, though some areas may be less smooth.\newline
- Ideas are well-organised and mostly easy to follow.\newline
\textbf{Example:} \emph{"Nurse 1's higher stress, as indicated by heart rate data, relates to her focus on Amy. Improving teamwork could help mitigate this issue."} \\
\cmidrule(lr){2-3}
& Level 3 (Good) &
- Displays basic organisation but may have occasional clarity issues.\newline
- Transitions between ideas may be abrupt or inconsistent.\newline
- Arguments are understandable but may lack depth.\newline
\textbf{Example:} \emph{"Nurse 1 worked more on Amy and had higher stress. Teamwork might have helped reduce her stress."} \\
\cmidrule(lr){2-3}
& Level 2 (Fair) &
- Ideas are poorly organised or disjointed.\newline
- Lacks clear transitions, making it hard to follow the argument.\newline
- Writing may be repetitive or off-topic.\newline
\textbf{Example:} \emph{"They worked alone a lot. Nurse 1's stress was high. Communication is important."} \\
\cmidrule(lr){2-3}
& Level 1 (Poor) &
- Lacks coherent structure; ideas appear random or jumbled.\newline
- Writing is confusing and difficult to understand.\newline
\textbf{Example:} \emph{"Nurse work on Amy. Stress high. Data shows tasks."} \\
\midrule

\multirow{5}{*}{Linguistic Quality}
& Level 5 (Excellent) &
- Writing is fluent with varied and precise vocabulary.\newline
- Virtually free of grammatical errors.\newline
- Complex ideas are expressed clearly and elegantly.\newline
\textbf{Example:} \emph{"The disproportionate workload on Nurse 1, evidenced by her solo focus on Amy and central communication role, resulted in elevated stress levels. A more equitable task distribution could mitigate stress and improve patient care."} \\
\cmidrule(lr){2-3}
& Level 4 (Very Good) &
- Generally clear and correct writing with minor errors.\newline
- Vocabulary is appropriate but may lack variety.\newline
- Ideas are communicated effectively.\newline
\textbf{Example:} \emph{"Nurse 1's higher stress relates to spending more time on Amy and communicating more. Sharing tasks might reduce her stress."} \\
\cmidrule(lr){2-3}
& Level 3 (Good) &
- Writing is understandable but contains noticeable errors.\newline
- Vocabulary and sentence structure are simple or repetitive.\newline
- Errors do not significantly impede understanding.\newline
\textbf{Example:} \emph{"Nurse 1 worked more with Amy and was more stressed than Nurse 2."} \\
\cmidrule(lr){2-3}
& Level 2 (Fair) &
- Frequent grammatical errors that may hinder comprehension.\newline
- Limited vocabulary; word choices may be inappropriate or incorrect.\newline
- Ideas are difficult to follow due to language issues.\newline
\textbf{Example:} \emph{"Stress Nurse 1 shows from Amy is cause by talks more and heart rate."} \\
\cmidrule(lr){2-3}
& Level 1 (Poor) &
- Writing is riddled with errors; meaning is often obscured.\newline
- Extremely limited vocabulary and flawed sentence structure.\newline
- Ideas are not effectively communicated.\newline
\textbf{Example:} \emph{"Nurse stress high. Work Amy. Data show."} \\
\midrule

\multirow{5}{*}{Critical Thinking}
& Level 5 (Excellent) &
- Demonstrates deep analysis and understanding of complexities.\newline
- Addresses possible counterarguments or alternative interpretations.\newline
- Provides well-reasoned, innovative solutions or recommendations.\newline
\textbf{Example:} \emph{"While prioritising Amy is essential, Nurse 1's elevated stress (111 BPM) and communication burden indicate an imbalance. Introducing a shared workload model could distribute tasks more evenly, potentially improving team performance and reducing individual stress."} \\
\cmidrule(lr){2-3}
& Level 4 (Very Good) &
- Analyses data carefully with awareness of underlying complexities.\newline
- Offers logical conclusions and considers some implications.\newline
- Suggestions are practical and grounded in the data.\newline
\textbf{Example:} \emph{"Nurse 1's stress suggests that although focusing on Amy is appropriate, collaborating more with colleagues might alleviate stress."} \\
\cmidrule(lr){2-3}
& Level 3 (Good) &
- Provides basic analysis but does not delve into complexities.\newline
- Conclusions are valid but may be obvious or lack depth.\newline
- Limited consideration of broader implications or solutions.\newline
\textbf{Example:} \emph{"Because Nurse 1 worked more on Amy, she was more stressed."} \\
\cmidrule(lr){2-3}
& Level 2 (Fair) &
- Offers superficial analysis with little critical engagement.\newline
- Conclusions may be unsupported or overly simplistic.\newline
\textbf{Example:} \emph{"Nurse 1 was stressed because she focused on Amy."} \\
\cmidrule(lr){2-3}
& Level 1 (Poor) &
- Lacks critical analysis; may misinterpret data.\newline
- Conclusions are incorrect or irrelevant.\newline
\textbf{Example:} \emph{"Nurse 1 stressed due to work on Amy. Nurse 2 worked less."} \\
\end{longtable}
\end{footnotesize}

\section{Scaffolding Prompt}
\label{sec:scaffold}

\begin{vizbox}{\new{Intervention Logic and Feedback Rules}}
\begin{itemize}
  \item \new{Track two state variables: the visualisation currently displayed and the guiding questions already covered for that visualisation.}
  \item \new{When a visualisation is first displayed, initiate with its one-sentence description and the first unanswered guiding question.}
  \item \new{After each learner response, compare the interpretation with the relevant ``knowledge to convey'' statement. Confirm an evidence-aligned interpretation, clarify omitted visual cues or relationships, and correct claims that contradict the displayed evidence. Do not draft the final paragraph for the learner.}
  \item \new{Advance to the next unanswered question after feedback. Do not repeat a guiding question and do not ask about a visualisation that is not currently displayed.}
  \item \new{If a response is procedural or off-task, briefly redirect attention to the current visual and continue the prescribed sequence.}
  \item \new{After all questions for one visualisation are completed, instruct the learner to navigate to the next visualisation. After all three are completed, issue the final synthesis prompt.}
\end{itemize}
\end{vizbox}

\begin{vizbox}
{General Guidelines}
\begin{itemize}
  \item Start with a brief, one-sentence description per visualisation.
  \item Only display and ask questions relevant to the visualisation currently shown.
  \item Do not ask the same question multiple times.
  \item After covering all guiding questions for a visualisation, prompt participants to move to the next by clicking the right arrow.
  \item Cover all guiding questions in order for each visualisation:
  \begin{itemize}
    \item Bar Chart: 1 guiding question
    \item Communication Network: 2 guiding questions
    \item Ward Map: 3 guiding questions
  \end{itemize}
  \item After finishing all visualisations, provide clear instructions for participants to complete the cohesive paragraph activity (see final prompt).
\end{itemize}
\end{vizbox}

\subsubsection*{Visualisations and Guiding Questions}

\begin{vizbox}{1. Bar Chart}
\descpt{``This bar chart shows the proportion of time the two nurses spent on different tasks.''}

\guiding
\begin{enumerate}
  \item What do you notice about the proportion of time spent on tasks for Amy compared to other tasks? What does this indicate about task prioritisation?
\end{enumerate}

\knowledge\; Make sure the participant understands how different proportions reveal each nurse’s prioritisation of their responsibilities.
\end{vizbox}

\begin{vizbox}{2. Communication Network}
\descpt{``This visualisation shows verbal communication interactions among each student and their communication with doctor, relative, and patient manikins. Arrows show the direction of communication, while arrow thickness indicates communication duration.''}

\guiding
\begin{enumerate}
  \item Who are the main communicators in the network, and what does the thickness of the arrows tell us?
  \item How often did the nurses communicate with each other compared to their communication with the doctor, patient, and relative?
\end{enumerate}

\knowledge\; Ensure the participant grasps which individuals are central communicators, what arrow thickness implies about communication duration, and differences between nurse-nurse and nurse-others interactions.
\end{vizbox}

\begin{vizbox}{3. Ward Map}
\descpt{``This Ward Map displays verbal and spatial distribution of each student during the simulation. More saturated colour shows frequent verbal communication, and hexagon locations indicate spatial movements. Peak heart rate for each student is also shown as heart symbols on the map.''}

\guiding
\begin{enumerate}
  \item Which areas of the ward did the nurses spend the most time in, and how does this reflect their task prioritisation?
  \item How can the colour intensity provide insights about nurses’ verbal communication patterns?
  \item What does peak heart rate suggest about nurses’ stress levels in different ward areas?
\end{enumerate}

\knowledge\; Make sure participants clearly understand the relationship between task prioritisation, verbal exchanges as indicated by saturated colours, and areas of high stress (heart rate).
\end{vizbox}

\begin{vizbox}{Final Activity (after all visualisations)}
When all visualisations and guiding questions are completed, instruct participants explicitly:

\begin{quote}
``Now, in the bottom-left space provided, please write a cohesive paragraph describing how the two nurses managed the primary patient (Amy) while attending to other beds, focusing on how they prioritised tasks, communicated verbally, and experienced stress as evidenced by the visualisations you’ve explored.''
\end{quote}
\end{vizbox}

\bibliographystyle{cas-model2-names}
\bibliography{0_reference}

\end{document}